\documentclass[aps, prc, twocolumn, floatfix, nofootinbib, preprintnumbers, superscriptaddress, amsmath, amssymb, longbibliography]{revtex4-2}

\usepackage{graphicx}
\usepackage{bbm}
\usepackage{slashed}
\usepackage{dcolumn}
\usepackage{siunitx}
\usepackage{multirow}
\usepackage{hhline}
\usepackage{enumitem}
\usepackage{mathrsfs}
\usepackage{xcolor}

\graphicspath{{Figures/}}

\usepackage{array}
\usepackage{ulem}

\definecolor{bobcatgreen}{rgb}{0.05,0.31,0.25}

\begin{document}

\def\lsim{\mathrel{\rlap{\lower4pt\hbox{\hskip1pt$\sim$}}
    \raise1pt\hbox{$<$}}}         %less than or approx. symbol
\def\gsim{\mathrel {\rlap{\lower4pt\hbox{\hskip1pt$\sim$}}
    \raise1pt\hbox{$>$}}}         %greater than or approx. symbol
    
\title{An effective field theory approach to fermionic rotational bands in nuclei}

\author{I.~K.~Alnamlah}
\email{ia151916@ohio.edu}
\affiliation{Institute of Nuclear and Particle Physics and Department of Physics and Astronomy, Ohio University, Athens, OH 45701,USA}
\affiliation{Department of Physics and Astronomy, King Saud University, Riyadh 11451, Saudi Arabia}

\author{E.~A.~Coello P\'erez}
\email{coelloperez1@llnl.gov}
\affiliation{Lawrence Livermore National Laboratory, Livermore, CA 94550, USA}
\affiliation{Institut f\"ur Kernphysik, Technische Universit\"at Darmstadt, 64289 Darmstadt, Germany}
\affiliation{ExtreMe Matter Institute EMMI, GSI Helmholtzzentrum f{\"u}r Schwerionenforschung GmbH, 64291 Darmstadt, Germany}

\author{D.~R. Phillips}
\email[]{phillid1@ohio.edu}
\affiliation{Institute of Nuclear and Particle Physics and Department of Physics and Astronomy, Ohio University, Athens, OH 45701,USA}
\affiliation{Institut f\"ur Kernphysik, Technische Universit\"at Darmstadt, 64289 Darmstadt, Germany}
\affiliation{ExtreMe Matter Institute EMMI, GSI Helmholtzzentrum f{\"u}r Schwerionenforschung GmbH, 64291 Darmstadt, Germany}

\date{\today}

\begin{abstract}
We extend an effective field theory developed to describe rotational bands in even-even nuclei to the odd-mass case. This organizes Bohr \& Mottelson's treatment of a particle coupled to a rotor as a model-independent expansion in powers of the angular velocity of the overall system. We carry out this expansion up to fourth order in the angular velocity and present results for $^{99}$Tc, ${}^{159}$Dy, ${}^{167, 169}$Er, ${}^{167, 169}$Tm, ${}^{183}$W, ${}^{235}$U and ${}^{239}$Pu. In each case,  the accuracy and breakdown scale of the effective field theory can be understood based on the single-particle and vibrational energy scales in that nucleus.
\end{abstract}

\maketitle

\section{Introduction}
For many even-even nuclei, the rotor model provides a good description of the energies of their low-lying states and of the transitions between them~\cite{BohrMottelson}. In this model, the nucleus is pictured as an axially symmetric quantum-mechanical rotor, whose eigenenergies are proportional to $I(I+1)$, with $I$ the spin of the nuclear state.
This picture can be extended to neighboring odd-mass nuclei by coupling a fermion to the rotor. This ``particle-rotor" approach can be quite successful in describing the low-lying spectra and transitions. However, it also works markedly better in some nuclei than in others. The classic text by Bohr and Mottelson provided an extensive summary of the successes and challenges of such a picture already fifty years ago~\cite{BohrMottelson}.

In this paper we re-cast the particle-rotor model as a systematic Effective Field Theory (EFT), building on the successful and systematic description of even-even systems as rotors by Coello P\'erez and Papenbrock~\cite{Papenbrock:2010yg,CoelloPerez:2015}. In that EFT the rotor degree of freedom is its angular velocity, $\vec{v}$, and the Lagrangian is organized in powers of this quantity. Here we develop an EFT that also includes the fermion's position and spin as degrees of freedom. There is similar recent work on a particle-rotor EFT by Papenbrock and Weidenm\"uller~\cite{Papenbrock:2020zhh}.
EFTs are systematic expansions for observables as they are organized in powers of a small parameter. In our particle-rotor EFT the small parameter is the angular velocity of the overall system. The EFT will thus be suitable for nuclei in which the energy associated with the rotor degree of freedom, $E_{\rm rot}$, is smaller than the energy required to excite the fermion to a new quantum state, $E_{\rm sp}$, or the energy at which the rotor ceases to be rigid, $E_{\rm vib}$. The EFT can then also be understood as an expansion in powers of $\epsilon_{\rm vib}\equiv E_{\rm rot}/E_{\rm vib}$ and $\epsilon_{\rm sp}\equiv E_{\rm rot}/E_{\rm sp}$.

At leading order (LO) in this expansion the rotor is rigid and the fermion attached to it is in a specific quantum state~\cite{Rowe}. The corresponding Lagrangian is the sum of that of a classical, axially-symmetric, rotor, parametrized by two Euler angles, and that for a fermion whose interaction with the rotor is governed by a specific potential, $V$, in the corotating or intrinsic frame. This is the starting point of Bohr and Mottelson's particle-rotor model too. It yields a LO Hamiltonian
\begin{equation}
H^{(0)} = H_{\rm rot} + H_{\rm ferm},
\label{eq:LOH}
\end{equation}
where
\begin{equation}
%H_{\rm rot}=\frac{\vec{R}^2}{2 {\cal I}_0},
H_{\rm rot}=\frac{\vec{R}^2}{2 {\cal I}_0},
\quad
H_{\rm ferm}= T + V(\vec{r},\vec{s}),
\end{equation}
%with $\vec{R}={\cal I}_0 \vec{v}$ the rotor angular momentum and ${\cal I}_0$ its moment of inertia, while $\vec{r}$ is the single-particle coordinate and $\vec{s}$ the spin of the fermion of mass $m$.
with $\vec{R}={\cal I}_0 \vec{v}$ the rotor angular momentum and ${\cal I}_0$ its moment of inertia, while $\vec{r}$ is the single-particle coordinate, $T$ the corresponding kinetic energy, and $\vec{s}$ the spin of the fermion.
%with $H_{\rm ferm}=-\frac{\hbar^2 \nabla^2}{2 m} + V(\vec{r},\vec{s})$, where $\vec{r}$ is the single-particle coordinate of the fermion of mass $m$ and $\vec{s}$ is its spin. Meanwhile $H_{\rm rot}=\frac{\hbar^2 \vec{R}^2}{2 {\cal I}_0}$, with $\vec{R}={\cal I}_0 \vec{v}$ the rotor angular momentum and ${\cal I}_0$ its moment of inertia. At LO the 
The eigenstates of $\hat{H}^{(0)}$ are direct products of eigenstates of $\hat{H}_{\rm rot}$ and eigenstates of $\hat{H}_{\rm ferm}$, i.e., the single-particle orbitals~\cite{Rowe}. The spectrum of the odd-mass system is then a sequence of rotational bands, each built on a single-particle orbital. A particular band is labeled by the projection of the fermion's total angular momentum, $\vec{j}$, on the axis of the rotor, i.e., the component of $\vec{j}/\hbar$ in the $3$-direction in the intrinsic frame, typically denoted $K$.

Since the rotor is axially symmetric, $K$ continues to be a good quantum number even when corrections to the LO Lagrangian are considered, i.e., when the fermion's degrees of freedom become coupled to $\vec{v}$.
Indeed,  since $V$ is the potential energy of the fermion in a rotating frame, rotational invariance requires that it contains a term proportional to $\vec{v}$: the Coriolis force. In this work we allocate all such couplings between $\vec{v}$ and the fermionic degrees of freedom to an additional piece of $H$, $H_{\rm coup}$:
\begin{equation}
H = H_{\rm rot}(\vec{v}) + H_{\rm ferm}(\vec{r},\vec{s}) + H_{\rm coup}(\vec{v},\vec{r},\vec{s}).
\end{equation}

EFTs are also model independent. To achieve that we make no assumption about $V(\vec{r},\vec{s})$ or about $H_{\rm coup}$, other than that both are axially symmetric and the latter can be expanded in powers of $\vec{v}$. This is in contrast to the recent Ref.~\cite{Chen:2020qbf}, which also tackled the rotor-plus-fermion problem, but assumed that $V$ was a deformed harmonic oscillator potential, thereby adopting a model in which the fermion's single-particle states are Nillson-model orbitals.

Corrections induced by ``cranking" the general fermion potential $V(\vec{r},\vec{s})$, i.e., effects resulting from the fermion's interaction with a rotating core, then appear in $H_{\rm coup}$. The corresponding terms involve undetermined coefficients that are not related to $V(\vec{r},\vec{s})$ by rotational symmetry.
The leading piece of $H_{\rm coup}$ has the form of the Coriolis force and can be derived by defining a covariant derivative of the fermionic field~\cite{Papenbrock:2020zhh}. This would appear to fix the coefficient of this $O(v)$ part of $H_{\rm coup}$---as happens in the textbook treatment in Ref.~\cite{BohrMottelson}. However, in an EFT all operators consistent with the symmetries are permitted, and the same operator can also be induced by effects at the high scale $E_{\rm sp}$. Therefore the coupling in this NLO piece of the Hamlitonian, which we denote $H_{\rm coup}^{(1)}$, is not fixed. 
%At this level, $H^{(0)} + H_{\rm coup}^{(1)}$ is the same as the leading-order Hamiltonian derived by Papenbrock and Weidenm\"uller~\cite{Papenbrock:2020zhh}.
%That is to say, $H_{\rm coup}^{(1)}$ has the form of the Coriolis force, but with an arbitrary coefficient. 
The impact of $H_{\rm coup}^{(1)}$ on the system's energy levels can be computed in first-order perturbation theory. As is well-known, in first order the effect is non-zero only for $K=1/2$ bands, and represents the first correction to the ``adiabatic limit" in which the fermion orbits are aligned with the symmetry axis of the deformed core.  The high-energy dynamics is then summarized in the resulting formula for the energy levels of the odd-mass rotor by a matrix element of a fermionic operator. 

At next-to-next-to-leading order (N$^2$LO), both $H_{\rm rot}$ and $H_{\rm coup}$ receive further corrections of order $O(v^2)$. Corrections to the former are due to the nonrigidity of the rotor.
This also affects $H_{\rm coup}$, as interaction with the spinning core can produce excitation to single-particle states with energies of order $E_{\rm sp}$. Such effects must be parametrized by an effective operator of order $O(v^2)$ or higher (see Sec.~\ref{sec:effectiveoperators}). These pieces of $H_{\rm coup}^{(2)}$ renormalize the energy shift obtained from two insertions of $H_{\rm coup}^{(1)}$, i.e., it gives the high-energy part of the second-order corrections to the adiabatic limit. While the even-even system is a straightforward expansion in $\epsilon_{\rm vib}$ the odd-mass system's energy levels  show an interplay of expansions in $\epsilon_{\rm vib}$ and in $\epsilon_{\rm sp}$.

In this paper we carry out this joint expansion up to fourth order in the expansion parameter, thus computing the energy levels of a rotational band in the odd-even system up to accuracy $\left(E_{\rm rot}/E_{\rm high}\right)^3$, with $E_{\rm high}$ of order either $E_{\rm sp}$ or $E_{\rm vib}$. The Hamiltonian's expansion yields an expansion for energy levels in powers of the total angular momentum quantum number $I$. At each order in the expansion, new parameters appear and must be fit to data. 

A common criticism of such a calculation is that it lacks predictive power. But our EFT for rotational bands is systematic: at $n$th order it yields a correction to the energy of the states in a nuclear rotational band that scales in a definite way with the expansion parameter and with $I$:
\begin{equation}
(\Delta E)_{\rm N^nLO} \sim E_{\rm rot} \epsilon_{\rm sp}^{n-1} I^n.
\label{eq:improvement}
\end{equation}
The error in the resulting $n$th-order EFT energy-level formula then scales as $I^{n+1}$.
By analyzing the residuals of the EFT's prediction at each order with respect to data we will show that such systematic improvement is indeed present in our description~\cite{Lepage:1997cs}. Moreover, we will show that the residuals encode information on the breakdown scale of the EFT. In general the convergence of the EFT is at least as good as is expected according to Eq.~(\ref{eq:improvement})and an {\it a priori} estimate of the energy scales $E_{\rm rot}$, $E_{\rm sp}$, and $E_{\rm vib}$ in the even-even and odd-mass nuclei under consideration.

In the traditional rotor-model literature the Coriolis force that represents the $n=1$ correction in Eq.~(\ref{eq:improvement}) results in a ``decoupling parameter" appearing in the formula for the energies of states in the band. In the particle-rotor EFT the decoupling parameter is not computed from single-particle matrix elements (cf. Ref.~\cite{Chen:2020qbf}). Our goal is not a microscopic description of the rotor-fermion system. Instead we seek a description that captures the long-distance features of this system, and parametrizes its short-distance details in terms of coefficients that are fit to data. EFT helps us obtain this organized phenomenology because it is agnostic about the short-distance details and organizes the energies of levels in the band in terms of an expansion in a small parameter. We do {\it not} expand observables in powers of the fermion angular momentum $\vec{j}$ or its projection on the rotor axis, $K$. This means that in our approach there is not just a single low-energy constant in $H_{\rm coup}^{(1)}$, instead there is a string of fermion operators that can multiply the Coriolis operator structure, each having its own coefficient. This interpretation of the ``Coriolis operator" differs from that of Ref.~\cite{Papenbrock:2020zhh}, although the interpretational difference has no practical consequences for the energy-level formula.
The lack of power counting for operators built solely from the fermion's degrees of freedom means that although in principle it should be possible to connect our EFT to shell-model or {\it ab initio} calculations~\cite{Caprio:2019yxh} of rotational bands, this task will be complicated in practice because the parameters in the formula for the band energies actually represent matrix elements of arbitrary functions of $j_3$, $\vec{j}$, and $\vec{r}$, as explained further in Sec.~\ref{sec:higherorder} below.

The rest of the paper is structured as follows. In Sec.~\ref{sec:effectiveoperators} we make some general remarks about the ``integrating out" process that leads to the EFT of rotational bands developed here. This justifies the statements regarding operator suppression above. In Sec.~\ref{sec:LOL} we specify the degrees of freedom, write down the leading-order Lagrangian, $L_{\rm rot} + L_{\rm ferm}$ and obtain the constants of the motion. Section~\ref{sec:higherorder} discusses the allowed operators that can appear in $L_{\rm rot}$ and $L_{\rm coup}$ and the quantities by which each operator is suppressed, i.e., it develops the power counting for our EFT. Section~\ref{sec:formula} then uses the resulting expansion to derive the rotational band formula order-by-order in $I$, see Eq.~(\ref{eq:improvement}). Section~\ref{sec:applications} studies the extent to which the EFT describes rotational bands in ${}^{167,169}$Er, ${}^{167,169}$Tm, ${}^{239}$Pu, ${}^{235}$U, ${}^{159}$Dy, ${}^{99}$Tc, and ${}^{183}$W. In each case we show that the accuracy of the band formula, and the number of levels for which it gives a systematic treatment of the spectrum, can be tied to the scales $E_{\rm rot}$. $E_{\rm sp}$, and $E_{\rm vib}$ for that particular nucleus. Finally, Sec.~\ref{sec:conclusions} offers a summary and avenues for future work. 

\section{From $\boldsymbol{A}$ nucleons to an EFT of fermion rotational bands}

\label{sec:effectiveoperators}

{\it Ab initio} calculations that solve the quantum-mechanical many-body problem for a given nuclear Hamiltonian are now feasible for nuclei as large as ${}^{100}$Sn~\cite{Morris:2017vxi}. A recent no-core configuration interaction calculation of the Be isotope chain examined the emergence of rotational and shell-model degrees of freedom in these systems~\cite{Caprio:2019yxh}. In this section we elucidate the relationship between a calculation with $A+1$ nucleonic degrees of freedom and the rotor-plus-fermion EFT developed in this paper. 

To get from one to the other we first imagine that we can solve the $A$-body problem and determine the spectrum of the corresponding Hamiltonian $H_A$. The solutions of 
\begin{equation}
H_A |\phi_{R;n} \rangle=E_{R;n} |\phi_{R;n} \rangle
\end{equation}
form a complete set of states for the $A$ particles that make up the rotor. We label them by their total angular momentum and by another index $n$ that allows us to enumerate states of the same $R$. We divide those states into two groups: ones in the lowest-lying rotational band and states involving excitations that are predominantly of vibrational or single-particle character. The rotational band then forms a space ${\cal P}$, that will be included in our EFT, while the higher-energy states form a complementary space ${\cal Q}$. The gap between ${\cal P}$ and ${\cal Q}$ is assumed to be of order $E_{\rm vib}$.

We now want to consider the interaction of the $A+1$th nucleon with the other $A$ nucleons. This can be done by computing the optical potential if the nucleon has positive energy~\cite{Rotureau:2016jpf} or via a state-dependent effective potential. We take:
\begin{equation}
-\frac{\hbar^2 \nabla^2}{2m} \delta_{R'R} + V_{{\rm eff}; R'R}(\vec{r},\vec{s})=\langle \phi_{R';1}|H_{{\rm eff};A+1}(E)|\phi_{R;1} \rangle,
\label{eq:Hsp}
\end{equation}
to define the effective single-particle Hamiltonian in the situation where the rotor transitions from a state with total angular momentum $R$ to one with angular momentum $R'$ in the ${\cal P}$ part of the space. The $H_{{\rm eff};A+1}(E)$ whose matrix element appears on the right-hand side is the effective Hamiltonian that results from integrating out the effects of rotor states in the ${\cal Q}$ space. It is equal to $H_{A+1}$ plus corrections suppressed by powers of $E/E_{\rm vib}$.

The hypothesis of the rotor-fermion picture is that in the intrinsic frame of the rotor the effective potential is the same for all states in the subspace ${\cal P}$, since those states are related to one another by rotations. Under this hypothesis the Hamiltonian defined by Eq.~(\ref{eq:Hsp}) can be taken to be $H_{\rm sp}(\vec{r},\vec{s})$: it depends only on the last nucleon's spin and position. However this is only true to the extent that the rotor is rigid, i.e., all the states in the space ${\cal P}$ are generated by rotations of the ground state of the rotor. Corrections to the picture then appears as a series in $\epsilon_{\rm vib}$.

In the intrinsic frame $H_{\rm sp}$  generates a set of single-particle states
\begin{equation}
H_{\rm sp} |\psi_{K} \rangle =E_{K} |\psi_{K} \rangle
\end{equation}
that are labeled by their spin projection on the intrinsic 3-axis, $K$. Since in that frame the rotor is not moving this spectrum has a ground state that is separated from all other states in the spectrum by an energy $\sim E_{\rm sp}$. 

We now focus on just that lowest state, sometimes called ``the bandhead". The particle-rotor EFT is based on the picture that the low-energy eigenstates of the $A+1$-body problem are, at LO in the EFT expansion, product states
\begin{equation}
|\psi_{K} \rangle |\phi_A \rangle,
\end{equation}
where $|\phi_A \rangle \in {\cal P}$. These states form a rotational band, all built on the band-head $K$, where the fermion is to be thought of as in a particular single-particle state, while the core occupies one---or a superposition of several---of the rotational states that make up its ground-state band. 

The EFT Hamiltonian for this rotational band is obtained from the single-particle Hamiltonian by integrating out all the single-particle states other than $K$. It therefore differs from $H_{\rm sp}$ by operators that are suppressed by $E/E_{\rm sp}$. In the rest of this section we determine the operators that can appear in this EFT Hamiltonian.

\subsection{Allowed operators}

In Ref.~\cite{CoelloPerez:2015} the Lagrangian, and hence the Hamiltonian, could not depend on $\hat{e}_3$, the rotor axis (the 3-axis in the body-fixed/intrinsic frame), because of spontaneous symmetry breaking. However, here the rotor is interacting with the fermion and dependence on $\hat{e}_3$ is permitted in $H_{\rm coup}$.
The rotor operators that can appear in $H_{\rm coup}$ in the EFT then have the parity and time-reversal properties listed in Table~\ref{tab:fermionops}. Meanwhile we take the fermionic operators appearing in $H_{\rm coup}$ to be its co-ordinate $\vec{r}$ and its total angular momentum $\vec{j}$. Since $\vec{r}$ can be transformed to $\{r,\vec{l}\}$ with $\vec{l}$ the angular momentum vector, and $\vec{j}=\vec{l} + \vec{s}$ the two vectors $\vec{r}$ and $\vec{s}$ are sufficient to completely describe the fermion's state. Their discrete-symmetry properties are then shown in Table~\ref{tab:fermionops}.

\begin{table}[htb]
\begin{tabular}{|l|c|c||l|c|c|}
\hline
Operator & P & T & Operator & P & T \\
\hline
$\vec{v}$ & + & - & $\vec{j}$ & + & - \\
$\hat{e}_3$ & - & + & $\vec{r}$ & - & + \\
%\hhline{|=|=|=|}
%$\vec{j}$ & + & - \\
%$\vec{r}$ & - & + \\
\hline
\end{tabular}
\caption{Operators describing the state of the rotor (left  block) and fermion (right block), together with their properties under parity and time reversal. Note that $\vec{v}$ is an angular velocity, which explains the otherwise peculiar looking parity assignment.}
\label{tab:fermionops}
\end{table}

We can also build even-rank tensors of the parity-mixed dot products, e.g.
\begin{equation}
\left(\hat{e}_3 \cdot \vec{j}\right)^{2}, \quad \left(\vec{v} \cdot \vec{r}\right)^{2} - \frac{1}{3} \vec{v}^{\ 2} \vec{r}^{\ 2}, \quad \vec{v}^{\ 2} \vec{r}^{\ 2}.
\end{equation}
The first operator in this list can be absorbed into $L_{\rm ferm}$ since $\hat{e}_3 \cdot \vec{j}=K$, the projection of the fermion spin on the rotor axis. The other operators listed cannot be eliminated in this way, and will appear in $L_{\rm coup}^{(2)}$, the Lagrangian that produces the piece of $H_{\rm coup}$ that is second order in $\vec{v}$.

Note that $\hat{e}_3$ can change the $R$ quantum number of the rotor. 
Symmetry under reflection in the rotor's central plane (${\cal R}$ symmetry~\cite{BohrMottelson}) guarantees that only even powers of $\hat{e}_3 \cdot \vec{r}$ can appear. Such operators then only change the rotor $R$ by 2, i.e., move from one rotor state ($0^+$, $2^+$, etc.) to another.

\subsection{Suppression by powers of $\boldsymbol{\vec{v}}$---and nothing else}

Permitted operators that couple rotor and fermionic degrees of freedom, and are not already accounted for in the fermionic potential are then, up to second order in $v$:
\begin{eqnarray}
O_{1}&=&\vec v \cdot \vec{j}; \nonumber\\
O_{2a}&=& \vec{v}^2 \vec{j}^2; \nonumber\\
O_{2b}&=&(\vec{v} \cdot \vec{j})^2 - \frac{1}{3} \vec{v}^2 \vec{j}^2; \nonumber\\
O_{2c}&=&\vec{v}^2 \vec{r}^2; \nonumber\\
O_{2d}&=&(\vec{v} \cdot \vec{r})^2 - \frac{1}{3} \vec{v}^2 \vec{r}^2.
\label{eq:operatorlist}
\end{eqnarray}
However, there is no reason for the expectation value of $\vec{j}^2$ in the fermionic state to be small. Indeed, we expect it to be a number of order one. Furthermore, the operator $\vec{r}^2$ should generate an expectation value of order $\hbar^2/(\sqrt{2 \mu E_{\rm sp}})^2$ since, by the uncertainty principle, the fermionic wave function should have this spatial extent. This means, then, that powers of $\vec{j}^2$ and $\vec{r}^2$ are not suppressed. The only expansion we have, then, is the one in powers of the rotor velocity $\vec{v}$. This fact is not apparent in the Lagrangians developed in Ref.~\cite{Papenbrock:2020zhh}.

In fact, each operator in the list above can be multiplied by an arbitrary function of the scalar (and P- and T-even) quantities $\vec{j}^2$, $(\vec{j} \cdot \hat{e}_3)^2$, $\vec{r}^2$, $(\vec{r} \cdot \vec{e}_3)^2$, and $(\vec{j} \cdot \vec{r})^2$. As long as we are concerned only with the fermionic matrix element for a specific single-particle state $|\psi_{K} \rangle$ this doesn't matter: it just means that the coefficient of the operator with a particular power of $v$ and a particular tensor structure is a matrix element of an arbitrary function of fermionic operators with the appropriate symmetry properties. But, since we don't know the fermionic wave function anyway, this additional ignorance regarding the fermionic operator has no practical consequence.
Unlike Bohr and Mottelson or Ref.~\cite{Chen:2020qbf} we do not try to compute the matrix elements of the fermionic operators in Eq.~(\ref{eq:operatorlist}) by assuming a particular description of single-particle states. Instead we fit them to data.

\section{Degrees of freedom and leading-order Lagrangian}

\label{sec:LOL}

\subsection{Parametrizing the rotor}

In this section we review the parametrization of the rotor introduced in Ref.~\cite{Papenbrock:2010yg} and used in Ref.~\cite{CoelloPerez:2015}.
We start with the transformation properties of the elements $g\equiv g(\alpha,\beta)$ under SO(3)/SO(2) rotations, as they parametrize the orientation of the rotor. Indeed, the rotation
\begin{equation}
g^{-1} = e ^{i \beta J_2}e ^{i \alpha J_3},
\end{equation}
where $\vec{J}$ is the generator of rotations,
aligns the laboratory or extrinsic reference frame with the co-rotating or intrinsic reference frame. In the latter, the $3$-axis coincides with the symmetry axis of the rotor.
The dynamics of the rotor are thus determined by the time derivative of $g$. For simplicity we study how $g^{-1}\partial_t g$ transforms under rotations. Being an element of the Lie algebra of SO(3), it can be written as
\begin{equation}
g^{-1}\partial_t g = iv_1J_1 + iv_2J_2 + iv_3J_3.
\end{equation}
Employing the Baker–Campbell–Hausdorff formula we write the components of $\vec{v}$ in terms of the rotor's orientation angles and their time derivatives
\begin{equation}
\label{velocities1}
\begin{gathered}
v_1=\dot{\alpha }\sin{\beta},\qquad
v_2=-\dot{\beta},\qquad
v_3=-\dot{\alpha }\cos{\beta}.
\end{gathered}
\end{equation}
Under the rotation $r\equiv r(\varphi,\theta,\gamma)$, the element $g^{-1}\partial_t g$ transforms into~\cite{Papenbrock:2010yg}
\begin{equation}
g^{-1}\partial_t g \rightarrow \widetilde{g}^{-1}\partial_t\widetilde{g} = i\widetilde{v}_1J_1 + i\widetilde{v}_2J_2 + i\widetilde{v}_3J_3.
\end{equation}
where $\widetilde{g}\equiv g(\widetilde{\alpha},\widetilde{\beta})$ and the angles $\widetilde{\alpha}$, $\widetilde{\beta}$ and $\widetilde{\gamma}$ are complicated functions of both the orientation angles and those defining the rotation $r$. The components of the transformed element $\widetilde{\vec{v}}$ are
\begin{equation}
\begin{gathered}
\begin{pmatrix}\widetilde{v}_1\\\widetilde{v}_2\end{pmatrix} =
\begin{pmatrix}\cos{\widetilde{\gamma}}&&-\sin{\widetilde{\gamma}}\\\sin{\widetilde{\gamma}}&&\cos{\widetilde{\gamma}}\end{pmatrix}
\begin{pmatrix}v_1\\v_2\end{pmatrix},\qquad
\widetilde{v}_3 = v_3 + \dot{\tilde{\gamma}}.
\end{gathered}
\end{equation}
These equations show that under an SO(3) rotation $v_1$ and $v_2$ transform as the $x$ and $y$ components of a vector would under the rotation $\widetilde{h}\equiv h(\widetilde{\gamma})=e^{-i\widetilde{\gamma}J_3}$, allowing us to easily write rotationally-invariant objects from these components. 

\subsection{Fermion representation}
The transformation properties of the fermion field were discerned in Ref.~\cite{Papenbrock:2010yg}. Let $\Phi$ represent the fermion field in the intrinsic reference frame. Then, the fermion field in the extrinsic frame, $\Psi$, can be written as
\begin{equation}
\Psi=g\Phi.
\end{equation}
From this expression, it can be shown that under rotations, the intrinsic fermion field transforms as 
\begin{equation}
    \Phi \rightarrow \widetilde{\Phi} = \widetilde{h}\Phi,
\end{equation}
%\begin{gather}
%\Psi \rightarrow \widetilde{\Psi} \equiv r \Psi = r g \Phi = \widetilde{g} \widetilde{h} \Phi = \widetilde{g}\widetilde{\Phi},
%\end{gather}
allowing us to write rotationally-invariant objects from the intrinsic fermion field with ease. Similarly, if the covariant derivative is defined as
\begin{equation}
D_t\equiv\partial_t + i v_3J_3 =\partial_t - i \dot{\alpha}\cos{\beta} J_3,
\end{equation}
the covariant derivative of the intrinsic field, $D_t\Phi$, transforms as
\begin{equation}
    D_t\Phi \rightarrow\widetilde{D}_t\widetilde{\Phi} = \widetilde{h}D_t\Phi.
\end{equation}

We can parametrize $\Phi$ in terms of the angles $\gamma$ and $\theta$ specifying the orientation of the fermion's total angular momentum in the intrinsic frame. For example, a fermion in a spin-half orbital would take the form
\begin{align}
     \Phi=\phi e^{-i\gamma J_3}e^{-i\theta J_2} \begin{pmatrix}1\\0\end{pmatrix},
\label{spin1/2a}
\end{align}
where $\phi$ is the solution to the part of the rotor's potential that does not change under rotations. This means that $\phi$ carries the radial dependence i.e., $\phi(r)$. 
We note that rotations around the symmetry axis are arbitrary for an axially-symmetric system. So we choose to write the fermion in a different form that will be useful in what follows. We write
\begin{align}
     \Phi= e^{-i\gamma {J}_3}\xi,
\end{align}
where $\xi$ is the fermion state in the intrinsic frame with the choice $\gamma=0$, i.e. 
\begin{equation}
\xi=\phi e^{-i\theta {J}_2} \begin{pmatrix}1\\0\end{pmatrix}.
\label{spin1/2b}
\end{equation}
We then rewrite the fermion in the extrinsic frame as
\begin{equation}
    \Psi=r(\alpha, \beta,\gamma)\xi.
\label{newparametrization}
\end{equation}
Even though we have shown the parametrization in equations (\ref{spin1/2a}) and (\ref{spin1/2b}) for a fermion in a spin-half orbital, equation (\ref{newparametrization}) is general. We also note that even though the choice of $\gamma$ is arbitrary $\dot{\gamma}$ isn't, since $\dot{\gamma}$ could couple to other degrees of freedom and we chose $\gamma$ to be part of the rotational degrees of freedom.

Note that this intrinsic frame constitutes an additional choice beyond that made at the start of this section: we exploit the symmetry of the rotor around the $3$-axis to choose an intrinsic frame in which the fermion spin vector lies in the $1$-$3$ plane. In this frame the components of the element $\vec{v}$ are obtained by studying the dynamics of the element $r^{-1}\partial_tr$:
\begin{equation*}
\begin{gathered}
v_1=\dot{\alpha }\sin{\beta}\cos{\gamma}-\dot{\beta}\sin{\gamma},\\
v_2=-\dot{\alpha }\sin{\beta}\sin{\gamma}-\dot{\beta}\cos{\gamma},\\
v_3=-\dot{\alpha }\cos{\beta}-\dot{\gamma}.
\end{gathered}
\end{equation*}

\subsection{Constructing the leading-order Lagrangian}

We separate the Lagrangian into a term involving the rotational degrees of freedom of the whole system (i.e. $\vec{v}$), a term involving the fermionic degrees of freedom, $\vec{r}$ and $\vec{s}$, and a term encoding the coupling between the fermion and the overall rotational motion
\begin{equation}
L=L_{\rm rot} + L_{\rm ferm} + L_{\rm coup}.
\end{equation}
Here
\begin{equation}
L_{\rm ferm}=\frac{1}{2} m \left(\frac{d \vec{r}}{dt}\right)^2 - V(\vec{r},\vec{s})
\end{equation}
 where $V(\vec{r},\vec{s})$ is the aforementioned single-particle effective potential.
Meanwhile $L_{\rm rot}$ is as discussed in Ref.~\cite{CoelloPerez:2015}, and is built out of powers of $\vec{v}$---or more specifically $v_{+1} v_{-1}$.

 A Lagrangian consisting of \(v_{+1}\) , \(v_{-1}\) , \(\xi\), and \(D_t\xi\) that is invariant under rotations of the subgroup \(SO(2)\) will be invariant under the full action of the group \( SO(3)\).
 We write the LO Lagrangian in the intrinsic frame as
\begin{equation}
\begin{aligned}
L_{\rm LO} =& C_0v_{+1}v_{-1} + \xi^\dagger i D_t \xi -\xi^\dagger H_{\rm int} \xi\\
=& \frac{C_0}{2}(\dot{\alpha}^2\sin^2{\beta}+\dot{\beta}^2)\\
&+q|\phi|^2\dot{\alpha} \cos{\beta} \cos{\theta}\\
&+q\dot{\gamma}|\phi|^2 \cos{\theta}+i\phi^*\dot{\phi}-\xi^\dagger H_{\rm int} \xi,
\end{aligned}
\label{eq:LOL}
\end{equation}
where $q$ is the total angular momentum of the fermion in the intrinsic frame.

The physics content of this Lagrangian is clearer if we compute the matrix elements of the fermion's angular momentum. Defining $j_k \equiv \xi^\dagger J_k\xi$ for $k=1,2,3$ we have
\begin{equation}
\begin{gathered}
j_1=q|\phi|^2 \sin{\theta},\\
j_2=0,\\
j_3=q|\phi|^2 \cos{\theta}.
\end{gathered}
\end{equation}
We also note that 
$i \xi^\dagger \partial_t \xi=i \phi^* \dot{\phi}$. 
The conjugate momenta for the co-ordinates $\phi$, $\alpha$, $\beta$, and $\gamma$ are then
\begin{equation}
\begin{gathered}
p_\phi = i\phi^*,\\
p_\alpha = C_0\dot{\alpha}\sin^2{\beta} + \cos{\beta}j_3,\\
p_\beta = C_0\dot{\beta},\\
p_\gamma=q|\phi|^2 \cos{\theta}.
\end{gathered}
\end{equation}
We note that $p_\gamma=\xi^\dagger J_3\xi$, so, as expected, the component of the total angular momentum along the rotor symmetry axis comes entirely from the fermion.
The LO Hamiltonian associated with the Lagrangian~(\ref{eq:LOL}) then takes the more transparent form
\begin{equation}
\begin{gathered}
H_{\rm LO} = \frac{1}{2C_0}\left[\left(\frac{p_\alpha-\cos{\beta}p_\gamma}{\sin{\beta}}\right)^2 + p_\beta^2 \right] +\xi^\dagger H_{int}\xi
\end{gathered}
\end{equation}

This Hamiltonian can be cast in an even simpler form if we rewrite it in terms of the total angular momentum of the system, $\vec{Q}$. The expressions for the components of this constant of motion in the extrinsic frame are (see Appendix~D of Ref.~\cite{Papenbrock:2010yg} for details)
\begin{equation}
\begin{gathered}
Q_x=-p_{\alpha}\cot{\beta}\cos{\alpha}-p_{\beta}\sin{\alpha}+p_\gamma \frac{\cos{\alpha}}{\sin{\beta}},\\
Q_y=-p_{\alpha}\cot{\beta}\sin{\alpha}+p_{\beta}\cos{\alpha}+p_\gamma \frac{\sin{\alpha}}{\sin{\beta}},\\
Q_z=p_{\alpha}.
\end{gathered}
\end{equation}
The expression for the intrinsic components $Q_1$, $Q_2$ and $Q_3$ can be obtained from the ones above by means of the rotation $r^{-1}$. In terms of the square of the total angular momentum,
\begin{equation}
\begin{gathered}
\vec{Q}^2=p_\beta^2+\frac{1}{\sin^2{\beta}}(p_{\alpha}-\cos{\beta}p_\gamma )^2+p_\gamma ^2
\end{gathered}
\end{equation}
the LO Hamiltonian can be written as
\begin{equation}
\begin{aligned}
H_{\rm LO} = \frac{1}{2C_0}(\vec{Q}^2-p_\gamma^2)+\xi^\dagger H_{int}\xi
%=& \frac{1}{2C_0}R^2+\Phi^\dagger H_{int}\Phi
\end{aligned}
\label{lohamiltonian}
\end{equation}

\section{Higher-order terms}

\label{sec:higherorder}

As discussed in section~\ref{sec:effectiveoperators}, the LO Lagrangian~(\ref{eq:LOL}) can be systematically improved by including permitted operators coupling rotor and fermion degrees of freedom with increasing powers of the low-energy operator $v$. These operators effectively account for the interaction between the rotor and the fermion and the nonrigidity of the former. 

The order-by-order construction of the effective Hamiltonian is achieved employing Fukuda's inversion method to expand the generalized velocities of the rotor, $\dot{x}\in\{\dot{\alpha},\dot{\beta}\}$, in power series of the dual expansion parameters $\epsilon_{\rm vib},\epsilon_{\rm sp}$:
\begin{equation}
\dot{x}=\sum_m\dot{x}^{(m)},
\end{equation}
where $\dot{x}^{(m)}\sim\dot{x}^{(0)}\epsilon^m$. We remind the reader that in this work we do not attempt an EFT expansion for the single-particle potential $V_{\rm sp}(\vec{r},\vec{s})$. 
Furthermore, since we have no power counting for fermionic operators any combination of them permitted by symmetries can multiply an operator of a given order in $v$.

Nevertheless, if we consider the leading-order Lagrangian for the rotor then, if we write the generalized velocities as expansions in powers of $(\epsilon_{\rm sp},\epsilon_{\rm vib})$, it can be symbolically written as
\begin{equation}
\begin{aligned}
L_{\rm LO} =& C_0 (v^{(0)} + v^{(1)} + \ldots)_{+1}(v^{(0)} + v^{(1)} + \ldots)_{-1} \\
=& C_0 v^{(0)}_{+1} v^{(0)}_{-1} + C_0 v^{(0)}_{+1} v^{(1)}_{-1} + C_0 v^{(1)}_{-1} v^{(0)}_{+1} +  \ldots \, .
\end{aligned}
\end{equation}
The second and third terms in the second line here are $(\epsilon_{\rm sp},\epsilon_{\rm vib})$ times smaller than the first one, and must be accounted for when computing the $O(v)$ piece of the Hamiltonian. The terms in $\ldots$ are suppressed by additional powers of the small parameter and are included in corrections to $H$ at $O(v^2)$ and beyond. 

\subsection{Leading rotor-fermion coupling}

At lowest order in $\vec{v}$, the only relevant term correcting the LO Lagrangian is
\begin{equation}
\Delta L_{\rm NLO} = C_1 (v_{+1} j_{-1} + v_{-1} j_{+1}).
\label{dNLOL}
\end{equation}
The next-to-leading order (NLO) correction to the Hamiltonian can be written as
\begin{equation}
\begin{aligned}
\Delta H_{\rm NLO} =& p_{\alpha}\dot{\alpha}^{(1)} + p_{\beta}\dot{\beta}^{(1)} -
L_{\rm LO}^{(1)}
- \Delta L_{\rm NLO}^{(1)}.
\end{aligned}
\end{equation}
Notice that this correction includes a contribution from the LO part of the Lagrangian, as discussed above. 
Inserting the expressions for the components of the generalized velocities in the above equation yields
\begin{equation}
\begin{aligned}
\Delta H_{\rm NLO} = \frac{C_1}{C_0} \left(j_{+1}Q_{-1} + j_{-1}Q_{+1}\right),
\end{aligned}
\label{eq:dNLOH}
\end{equation}
where we have defined $j_{\pm 1}=\xi^\dagger(J_1\pm i J_2)\xi/\sqrt{2}=q|\phi|^2 \sin{\theta}/\sqrt{2}$ and $Q_{\pm 1}=(Q_1\pm i Q_2)/\sqrt{2}$, and used the identity
\begin{equation}
\begin{aligned}
j_{+1}Q_{-1} +& j_{-1}Q_{+1}=q|\phi|^2\sin{\theta} \\
\times& \left(p_\beta  \sin{\gamma} - \frac{p_\alpha}{\sin\beta} \cos{\gamma} + \frac{\cos\beta}{\sin\beta} p_\gamma\cos{\gamma} \right).
\end{aligned}
\end{equation}

The operator structure of this correction is identical to that of the Coriolis term obtained by writing a rotationally-invariant Lagrangian for the particle-rotor system in the extrinsic frame
\begin{equation}
L_{\rm ext} = C_0 v_{+1}v_{-1} + \Psi^{\dagger} i\partial_t \Psi - \Psi^{\dagger} H_{\rm ext}^{(\rm f)} \Psi,
\end{equation}
and rotating it to the intrinsic one by means of the rotation $r$
\begin{equation}
\begin{aligned}
L_{\rm int} =& C_0 v_{+1}v_{-1} + \xi^{\dagger} r^{-1}i\partial_t r \xi - \xi^{\dagger} r^{-1}H_{\rm ext}^{(\rm f)}r \xi\\
=& C_0 v_{+1}v_{-1} - \vec{v}\cdot\vec{j} + \xi^{\dagger} i\partial_t \xi - \xi^{\dagger} H_{\rm int}^{(\rm f)} \xi,
\end{aligned}
\end{equation}
This intrinsic Lagrangian yields a Hamiltonian
\begin{equation}
\begin{aligned}
H_{\rm int} =& \frac{1}{2C_0} \left(\vec{Q} - \vec{j} \right)^2 + \xi^{\dagger} H_{\rm int}^{(\rm f)} \xi\\
=& \frac{1}{2C_0} \left[ \vec{Q}^2 - 2\left(j_{+1}Q_{-1} + j_{-1}Q_{+1}\right) \right] + \xi^{\dagger} \widetilde{H}_{\rm int}^{(\rm f)} \xi,
\end{aligned}
\end{equation}
with $\widetilde{H}_{\rm int}^{(\rm f)}\equiv H_{\rm int}^{(\rm f)} + \vec{j}^2 - 2j_3^2$. This is equivalent to the rotor-fermion model Hamiltonian of Bohr and Mottelson~\cite{BohrMottelson}, and similar to our NLO effective Hamiltonian, $H_{\rm NLO}=H_{\rm LO}+\Delta H_{\rm NLO}$. Notice, however, that the coefficient accompanying the ``Coriolis'' term stemming from Eq.~(\ref{eq:dNLOH}) is undetermined. In contrast the Coriolis term in the Bohr and Mottelson Hamiltonian is determined by the requirement of rotational invariance in the extrinsic frame---this is, after all, a classical-mechanics argument to this point.

Papenbrock and Weidenm\"uller point out that the fact that $C_1$ is not determined by symmetries and so is $\neq -1$ can be understood as a consequence of the presence of a gauge coupling that modifies the interaction of the fermion with the rotor velocity field that is generated by minimal subtitution~\cite{Papenbrock:2020zhh}. 
 In fact, the coefficient appearing in Eq.~(\ref{eq:dNLOH}) should not be understood as a number: any fermionic operator---$\vec{j}^2$, $(\vec{j} \cdot \hat{e}_3)^2$, $\vec{r}^{\, 2}$, $(\hat{e}_3 \cdot \vec{r})^{\, 2}$---can appear in it. Therefore the full correction to the Hamiltonian at NLO in our expansion in powers of $\epsilon_{\rm sp}$ is:
\begin{equation}
\Delta {H}_{\rm NLO} = f(\vec{j}^{\, 2},(\vec{j} \cdot \hat{e}_3)^2, \vec{r}^{\, 2}, (\hat{e}_3 \cdot \vec{r})^{\, 2}) (j_{+1} Q_{-1} + j_{-1} Q_{+1}),
\label{eq:dHNLOfull}
\end{equation}
where $f$ encodes an arbitrary string of fermionic operators. We will see below that the distinction between $f$ and $C_1/C_0$ is irrelevant as far as practical application of this EFT is concerned, since the matrix element of the quantum-mechanical operator $f(\ldots) \vec{j}$ in the fermionic state on which the band is built determines the size of the NLO effect in that band. 

\subsection{Corrections to the rotor-fermion coupling}
Operators involving more powers of $v$ improve the rotor-fermion interaction in Eq.~(\ref{dNLOL}). The $O(v^2)$ correction to this ``Coriolis'' term is
\begin{equation}
\Delta L_{\rm N^2LO} = \frac{C_2}{2} (v_{+1} j_{-1} +v_{-1} j_{+1})^2 + D_2(v_{+1} v_{-1})\vec{j}^2.
\end{equation}
Including this contribution to the effective Lagrangian and the second-order components of the generalized velocities in the effective Hamiltonian yields an N$^2$LO correction
\begin{equation}
\begin{aligned}
\Delta H_{\rm N^2LO} =& -\frac{C_2}{2C_0^2}(j_{+1}Q_{-1}+j_{-1}Q_{+1})^2\\
&- \frac{D_2}{2C_0^2}(\vec{Q}^2-j_3^2)\vec{j}^{\, 2}\\
&+ \frac{C_1^2}{C_0}(j_1^2+j_2^2).
\end{aligned}
\label{eq:dN2LOH}
\end{equation}
It is important to mention that our lack of knowledge of the details of the single-particle states makes it impossible to disentangle the matrix elements of the third term in this correction from those of $H_{\rm ferm}$. The effects of this term are thus taken into account already at LO. 

In a similar way, including the N$^3$LO contribution to the Lagrangian
\begin{equation}
\begin{aligned}
\Delta L_{\rm N^3LO} &= \frac{C_3}{3} (v_{+1} j_{-1} + v_{-1} j_{+1})^3 \\
&+ \frac{2D_3}{3}(v_{+1} v_{-1})(v_{+1} j_{-1} + v_{-1} j_{+1})
\end{aligned}
\end{equation}
and the third-order components of the generalized velocities, yields the N$^3$LO correction to the Hamiltonian
\begin{equation}
\begin{aligned}
\Delta H_{\rm N^3LO} &= \frac{C_3}{3C_0^3} (j_{+1}Q_{-1}+j_{-1}Q_{+1})^3\\
&+ \frac{D_3}{3C_0^3}(\vec{Q}^{\, 2}-j_3^2)(j_{+1}Q_{-1}+j_{-1}Q_{+1})\\
&- \frac{C_1}{C_0^2}[j_{+1}Q_{-1}+j_{-1}Q_{+1}][C_2(j_1^2+j_2^2)+D_2 {\vec j}^{\, 2}].
\label{dHN3LO}
\end{aligned}
\end{equation}
Again, the matrix elements of the third term cannot be disentangled from those of $\Delta H_{\rm NLO}$, and thus those effects are taken into account at that order. 

This pattern is generic: 
the relevant terms in the order $n$ correction to the Hamiltonian always come from $-\Delta L_{\rm N^nLO}(\dot{x}^{(0)})$. Terms in the $n$th-order piece of the Lagrangian, $\Delta L_{\rm N^nLO}$, that contain $(v_{+1}v_{-1})^n$ and $(v_{+1} j_{-1} +v_{-1} j_{+1})^n$ therefore translate into corrections to the Hamiltonian containing $(\vec{Q}^2-j_3^2)^n/2^nC_0^{2n}$ and $(j_{+1}Q_{-1}+j_{-1}Q_{+1})^n/C_0^n$, respectively.

It then follows that the fourth-order contribution to the Lagrangian,
\begin{equation}
\begin{aligned}
\Delta L_{\rm N^4LO} =& \frac{C_4}{4}(v_{+1} j_{-1} +v_{-1} j_{+1})^4\\
&+ \frac{D_4}{2}(v_{+1}v_{-1})(v_{+1}j_{-1}+v_{-1}j_{+1})^2,
\end{aligned}
\end{equation}
yields the correction to the Hamiltonian
\begin{equation}
\begin{aligned}
\Delta H_{\rm N^4LO} =& \frac{C_4}{4C_0^4}(j_{+1}Q_{-1}+j_{-1}Q_{+1})^4\\
&+ \frac{D_4}{4C_0^4}(\vec{Q}^{\, 2}-j_3^2)(j_{+1}Q_{-1}+j_{-1}Q_{+1})^2\\
&+ \ldots
\label{HN4LOa}
\end{aligned}
\end{equation}
where the dots stand for terms whose matrix elements cannot be disentangled from those of lower-order corrections.

\subsection{Corrections to the rotor Lagrangian}
Besides effectively accounting for the interaction between the nucleons in the rotor and the fermion, higher-order contributions to the Lagrangian account for other effects. As shown in Ref.~\cite{CoelloPerez:2015} adding a term of the form
\begin{equation}
\Delta L_{\rm rotor~subleading}= E_4 (v_{+1}v_{-1})^2
\end{equation}
to the effective Lagrangian improves the description of the ground-state rotational bands in even-even nuclei. Including this term in our Lagrangian yields the correction to the Hamiltonian
\begin{equation}
\Delta H_{\rm rotor~subleading} = \frac{E_4}{4C_0^4}(\vec{Q}^{\, 2}-j_3^2)^2.
\label{HN4LOb}
\end{equation}
This correction, however, is not suppressed in the same way as the operators discussed up until this point since it does not arise from integrating out fermionic states. In Ref.~\cite{CoelloPerez:2015} the suppression of this correction was established to be $(E_{\rm rot}/E_{\rm vib})^2=\epsilon_{\rm vib}^2$.
For most of the systems considered in this work $(E_{\rm rot}/E_{\rm vib})^2\lsim (E_{\rm rot}/E_{\rm high})^3$, and thus we can treat the correction~(\ref{HN4LOb}) as part of the N$^4$LO correction to the Hamiltonian.

\section{Energy formula\\for rotational bands}
\label{sec:formula}

Now that we have an order-by-order expansion for the Hamiltonian of the rotor-fermion system in hand, we will use it to systematically compute the energies of states in rotational bands with definite $K$, seeking an expansion for those energies in powers of the same two expansion parameters used to organize the Hamiltonian in the previous section, i.e., $\epsilon_{\rm sp}$ and $\epsilon_{\rm rot}$.

\subsection{Leading order}   

Let us return to the expression for the LO Hamiltonian in Eq.~(\ref{lohamiltonian}). In what follows we take this classical function of the co-ordinates $\alpha$, $\beta$, $\gamma$, $\theta$ and $\phi$, and treat it as a quantum-mechanical operator with $\hat{Q}$ and $\hat{J}$ acting on the rotational and fermionic degrees of freedom, respectively. The eigenstates of the intrinsic single-particle Hamiltonian $\hat{H}_{\rm int}$, simultaneously eigenstates of $\hat{Q}_3$, are denoted by $\xi_K$
\begin{equation}
\hat{H}_{\rm int} \xi_K = E_K \xi_K.
\end{equation}
We note that since these states are eigenstates of $\hat{Q}_3$ they do not correspond to a definite $\theta$, i.e., the representation (\ref{spin1/2b}) does not apply from this point on. For the rotational portion of the Hamiltonian we choose the Euler angle ($\alpha$-$\beta$-$\gamma$) representation, for which the eigenstates of the rotor Hamiltonian $\hat{H}_{\rm rot}=(\hat{Q}^2-\hat{p}_\gamma^2)/2C_0$ are Wigner D-functions, denoted by $\mathscr{D}_{MK}^I\equiv \mathscr{D}_{MK}^I(\alpha,\beta,\gamma)$
\begin{equation}
\hat{H}_{\rm rot} \mathscr{D}_{MK}^I = \frac{\hbar^2 I(I+1)- \hbar^2 K^2}{2C_0} \mathscr{D}_{MK}^I.
\end{equation}

We focus on nuclei for which the band under study, defined by its value for $K$, is well separated from bands with $\hat{Q}_3=\hbar(K\pm 1)$. For these systems states in the band of interest can be described in terms of only one single-particle state $\xi_K$: mixing between it and $\xi_{K\pm 1}$ is a perturbative effect.
If the splitting between $E_K$ and $E_{K+1}$ (say) is ``accidentally" smaller than the typical single-particle energy scale $E_{\rm sp}$ (instead it is of order $E_{\rm rot}$) then both states must be considered as low-energy degrees of freedom in the rotational EFT. Significant inter-band mixing results. The consequences of this were discussed by Rowe for the case of ${}^{183}$W (cf. also Sec.~\ref{sec:W183} below)~\cite{Rowe} and by Papenbrock and Weidenm\"uller for ${}^{187}$Os~\cite{Papenbrock:2020zhh}.

But, in nuclei for which $|E_{K \pm 1} - E_K| \gg E_{\rm rot}$
product states
\begin{equation}
\left(\frac{2I+1}{8\pi^2}\right)^{1/2}\xi_K \mathscr{D}^I_{MK}
\label{eq:productstate}
\end{equation}
are eigenstates of $\hat{H}_{\rm LO}$ corresponding to eigenvalues 
\begin{equation}
\begin{aligned}
E_{\rm LO}(I,K) =& \langle KIM | \hat{H}_{\rm LO} | KIM \rangle\\
=& A_{\rm rot} I(I+1)+\widetilde{E}_K.
\end{aligned}
\label{eq:ELO}
\end{equation}
Here $A_{\rm rot}$ is the constant $\hbar^2/2C_0$ and $\widetilde{E}_K$ is the energy of the fermion in the intrinsic frame, $E_K$, minus $\hbar^2 K^2/2C_0$.
It is important to note that the Euler angles in the Wigner D-function describe the motion of the system as a whole, and not just that of the rotor to which the fermion is coupled. Correspondingly, $I$, $M$, and $K$ are the eigenvalues of $\hat{Q}^2$, $\hat{Q}_z$, and $\hat{Q}_3$, where $\hat{Q}$ is the total angular-momentum operator of the rotor-plus-fermion system. This system is then described by a wave function that is the product of the wave function of the fermion in the intrinsic frame and the wave function of the rotational motion of the system as a whole~\cite{BohrMottelson}.

Equation (\ref{eq:ELO}) is  the leading term in an adiabatic expansion for the energy of states in the $K$ rotational band. The adiabatic expansion is useful when ($\hbar$ times) 
the rotational frequency of the system is small compared to the excitation energies of the fermion in the rotor's potential. That states of the form (\ref{eq:productstate}) represent the wave function of the rotor-plus-fermion system at LO in such an expansion is emphasized by, e.g., Rowe in Ref.~\cite{Rowe}. 

The ${\cal R}$ symmetry of the system results in the state proportional to $\xi_{\bar{K}}\mathscr{D}^I_{M-K}$ having the same energy as the state~(\ref{eq:productstate}). ($\xi_{\bar K}$ is obtained from $\xi_K$ by applying the ${\cal R}$-parity operator, see Bohr and Mottelson, Eq.~(4-17).) 
It follows that the LO wave function of the rotor-fermion system is the one written in Bohr and Mottelson Eq.~(4A-5)
\begin{equation}
\Psi_{KIM}= \left(\frac{2I+1}{16\pi^2}\right)^{1/2}\left[\xi_K\mathscr{D}^I_{MK}+(-1)^{I+K}\xi_{\bar{K}}\mathscr{D}^I_{M-K} \right].
\label{wavefunctions}
\end{equation}

\subsection{Next-to-leading order}

The first-order correction to the adiabatic picture is generated by the NLO piece of the effective Hamiltonian. This term is linear in $v$ and couples the angular velocity of the system to the fermion’s degrees of freedom. It has a similar form to the well-known Coriolis coupling of classical mechanics. The expectation value of the NLO Hamiltonian, $\hat{H}_{\rm NLO}=\hat{H}_{\rm LO}+\Delta\hat{H}_{\rm NLO}$, for states in a $K$ band is
\begin{equation}
\begin{aligned}
E_{\rm NLO}&(I,K) = A_{\rm rot}I(I+1) + \widetilde{E}_K\\
& + A_1(-1)^{I+1/2}\left(I+\tfrac{1}{2}\right)\delta_K^{1/2}.
\end{aligned}
\end{equation}
Here we have absorbed all corrections to the fermion's energy into $\widetilde{E}_K$, and defined the LEC $A_1\equiv -a\hbar C_1/2C_0$, with $a=-\langle{K=1/2}|\sqrt{2}\hat{J}_{+1}|{\overline{K=1/2}}\rangle$ being Bohr and Mottelson's decoupling constant (notice the difference in the convention for $\hat{J}_{+1}$). This yields the expectation that the LEC $A_1$ is of order $A_{\rm rot}$ times the single-particle $J$. 
In this work we do not calculate $a$ since we are agnostic regarding the dynamics in $H_{\rm int}$. Instead we absorb this matrix element in the LEC $A_1$ and fit it to data.

The NLO correction to the energies is zero for all bands with $K \neq 1/2$ since $\Delta\hat{H}_{\rm NLO}$ changes $K$ by one unit. 
The last term in $E_{\rm NLO}(I,K)$ is sometimes called the signature term, and it causes staggering between adjacent states in $K=1/2$ bands. This staggering is clearly visible in experimental data. Thus, the addition of the NLO correction to the energies should improve the description of $K=1/2$ bands. 

We note that the Coriolis-like term in $H_{\rm NLO}$ can be treated in perturbation theory because we assume that the splitting between the $K=1/2$ and $K=3/2$ band (say) is large compared to the shift in energy induced by the NLO Hamiltonian. This provides a criterion for when this NLO term should be treated non-perturbatively. If the difference of band-head energies becomes of order $\Delta E_{\rm NLO}$ then $\Delta H_{\rm NLO}$ must be diagonalized in the basis of states $|\pm K \rangle$ and $|\pm (K+1) \rangle$. The result of this diagnoalization is worked out by Papenbrock and Weidenm\"uller in Ref.~\cite{Papenbrock:2020zhh} and then employed in ${}^{187}$Os. Here we restrict ourselves to situations were $E_{K+1} - E_K \gg \Delta E_{\rm NLO}$ and so perturbation theory is applicable. 

\subsection{Next-to-next-to-leading order}

At next-to-next-to-leading order (N$^2$LO) we should in principle consider two insertions of the Coriolis-like operator (\ref{eq:dHNLOfull}). This second-order correction accounts for virtual excitations of the fermion from, for example, a $K=1/2$ band to a $K=3/2$ band and back to the $K=1/2$ band. The calculation of these effects is discussed in Ref.~\cite{BohrMottelson}.  
However, in an EFT in which only one fermionic state is a low-energy degree of freedom there are no states to sum over in the second-order perturbation theory calculation. All such second-order effects are ``high-energy physics" and, as such, get subsumed into the operators that appear in the N$^2$LO Hamiltonian, $\hat{H}_{\rm N^2LO}=\hat{H}_{\rm NLO}+\Delta\hat{H}_{\rm N^2LO}$. In particular, the first term in Eq.~(\ref{eq:dN2LOH}) has the same operator structure as two insertions of the Coriolis-like operator. We calculate its contribution and that of the second term in Eq.~(\ref{eq:dN2LOH}) in Appendix \ref{app:2}, and rewrite the resulting N$^2$LO shift in the energy as a $K$-band dependent shift in the LEC $A_{\rm rot}$, i.e.:
\begin{equation}
\Delta E_{\rm N^2LO}(I,K)=\Delta A_K I(I+1),
\end{equation}
where
\begin{equation}
\Delta A_K \equiv -\frac{\hbar^2}{2C_0^2}\langle KIM|\left(C_2\hat{J}_{+1}\hat{J}_{-1} +  D_2\hat{J}^2\right)|KIM \rangle.
\end{equation}
Taking $C_2$ and $D_2$ to be of order $E_{\rm sp}^{-1}$ or $E_{\rm vib}^{-1}$ and $\hbar^2/C_0$ of order $E_{\rm rot}$ gives us an estimate for $\Delta A_K/A$ of order $\mathcal{O}(\epsilon_{\rm sp},\epsilon_{\rm vib})$. Meanwhile,  the contribution to the energy shift from the last term in the operator version of Eq.~(\ref{eq:dN2LOH}) can be absorbed into a redefinition of the energy $\widetilde{E}_K$ of the fermion, since the operator contributes only to $H_{\rm ferm}$. Hence, the energy formula up to N$^2$LO is
\begin{equation}
\begin{aligned}
E_{\rm N^2LO}&(I,K) = A_K I(I+1) + \tilde{E}_K\\ 
& + A_1(-1)^{I+1/2}(I+1/2)\delta_K^{1/2}.
\end{aligned}
\end{equation}
This appears to be the same as $E_{\rm NLO}$. However, $A_K$ includes $\Delta A$, which depends on fermionic matrix elements. This means we should fit $A_K$ to the odd-mass system. %, in which case we have $\Delta A_K = A_{\rm rot}-A^{odd}$.
We find that $A_K$ tends to be smaller than $A_{\rm rot}$ for ground-state bands. This is expected if $\Delta A_K$ is dominated by the piece $\sim C_2$, since second-order perturbations to a ground-state energy will result in a change in $A$ in this direction~\cite{BohrMottelson}. 

\begin{figure*}[t]
    \centering
    \includegraphics[width=\textwidth]{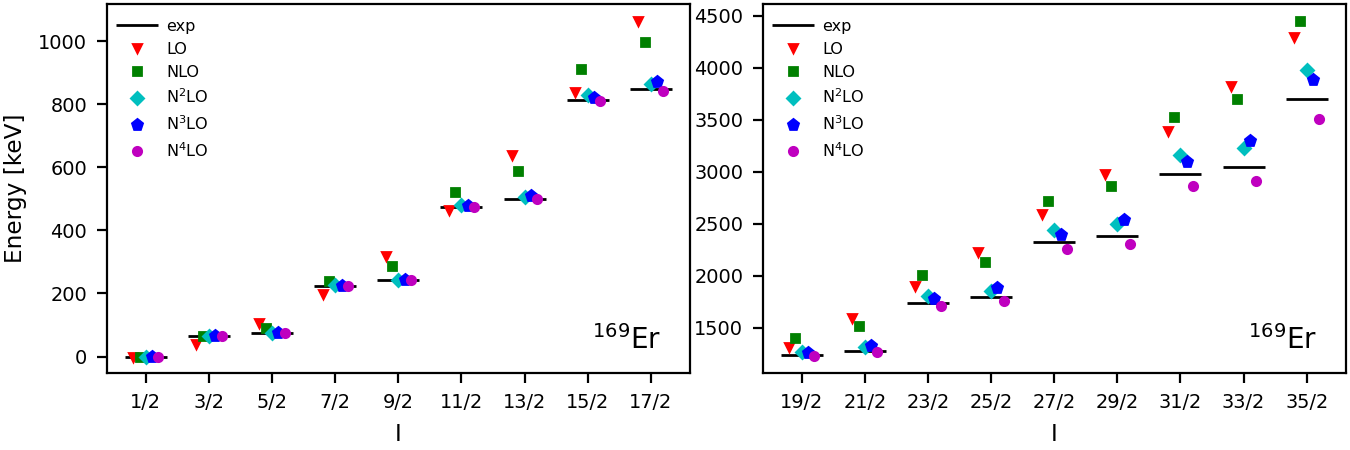}
    \caption{Calculated energies for the 1/2$^-$ ground-state rotational band in $^{169}$Er. The black line shows experimental values taken from the NNDC \cite{Baglin:2008hsa}. Red triangles, green squares, cyan diamonds, blue pentagons and magenta circles show calculated energies at LO, NLO, N$^2$LO, N$^3$LO, and N$^4$LO, respectively. The right panel is a continuation of the left panel with a different scale for the y-axis.}
    \label{fig:E_169Er}
\end{figure*}

\subsection{N$\boldsymbol{^3}$LO}

The pattern continues at N$^3$LO. The first operator in Eq.~(\ref{dHN3LO}) has the operator structure of three insertions of the Coriolis operator, but since no other states in the theory are dynamical, the sum over other states is replaced by an overall constant $C_3/3 C_0^3$. This operator permits the $K=3/2$ band to couple to itself thereby producing a signature term for the energies of $K=3/2$ states. Meanwhile, the Coriolis-like operator that gave rise to the signature term in $K=1/2$ bands at NLO is itself modified through multiplication by a factor of $\hat{Q}^2$. 
Thus, the energy formula at N$^3$LO takes the form
\begin{equation}
\begin{aligned}
E_{\rm N^3LO}&(I,K) = A_K I(I+1) + \tilde{E}_K\\ 
& + A_1(-1)^{I+1/2}(I+1/2)\delta_K^{1/2} \\
& \qquad + B_1 (-1)^{I+1/2}  I(I+1) \left(I+\tfrac{1}{2}\right) \delta_K^{1/2} \\
& \qquad + A_3 (-1)^{I+3/2} \left(I-\tfrac{1}{2}\right) \left(I+\tfrac{1}{2}\right) \left(I+\tfrac{3}{2}\right) \delta_K^{3/2}.
\end{aligned}
\end{equation}
Since these effects occur at N$^3$LO, we expect $B_1/A_{\rm rot}$ and $A_3/A_{\rm rot}$ to be of be of order $(\epsilon_{\rm sp},\epsilon_{\rm vib})^2$.

\subsection{N$\boldsymbol{^4}$LO}

Including the operators in Eqs.~(\ref{HN4LOa}) and (\ref{HN4LOb}) yields the N$^4$LO energy formula
\begin{equation}
\begin{aligned}
E_{\rm N^4LO}&(I,K) = A_K I(I+1) + \tilde{E}_K\\ 
& + A_1(-1)^{I+1/2}(I+1/2)\delta_K^{1/2} \\
& \qquad + B_1 (-1)^{I+1/2}  I(I+1) \left(I+\tfrac{1}{2}\right) \delta_K^{1/2} \\
& \qquad + A_3 (-1)^{I+3/2} \left(I-\tfrac{1}{2}\right) \left(I+\tfrac{1}{2}\right) \left(I+\tfrac{3}{2}\right) \delta_K^{3/2} \\
& \qquad\qquad + B_K I^2 (I+1)^2,
\end{aligned}
\end{equation}
where we defined $B_K\equiv B_{\rm rot}+\Delta B_K$, with $B_{\rm rot}$ and $\Delta B_K$ resulting from the contributions in Eqs.~(\ref{HN4LOb}) and~(\ref{HN4LOa}), respectively. We note that $B_{\rm rot} \sim E_{\rm rot}\epsilon_{\rm vib}^2$ while $\Delta B_K \sim E_{\rm rot}\epsilon_{\rm sp}^3$. As mentioned above, these two are of roughly the same size for the nuclei studied in this work. We therefore assign both effects to N$^4$LO in our EFT. The difference between $B_K$ and $B_{\rm rot}$ tends to be much larger (in fractional terms) than that between $A_K$ and $A_{\rm rot}$.

\section{Application}
\label{sec:applications}

We show the bandhead state properties, relevant energy scales, and ratios of low-energy constants (LECs) for the systems ${}^{99}$Tc, ${}^{159}$Dy, ${}^{167,169}$Er, ${}^{167,169}$Tm, ${}^{183}$W, ${}^{235}$U and ${}^{239}$Pu in Table \ref{table:1}. In the ``Energy Scales" part of the table $E_{\rm rot}$ is taken to be the energy of the first 2${^+}$ state in the ground-state rotational band of the rotor. The energy scale of vibration, $E_{\rm vib}$, is the energy of the first vibrational energy level of the rotor, and $E_{\rm sp}$ is the difference in energy between the energy of the specific band we are looking at and the next band that couples with it with $|\Delta K| \leq 1$. We see a clear trend that $\epsilon_{\rm vib}$ decreases with increasing mass. However, $\epsilon_{\rm sp}$ doesn't seem to have a clear trend. The results in the table for LEC ratios are discussed below, in Sec.~\ref{sec:LECs}.

Here we will, though, briefly describe the procedure we used to fit the LECs in the energy formula for $K=1/2$ bands.
At LO, the LECs $A_{\rm rot}$ and $\widetilde{E}_K$ are fitted to the energies of the 2$^+$ state in the ground-state rotational band of the rotor and the bandhead of the $K=1/2$ rotational band under consideration, respectively.
At NLO, the LECs $\widetilde{E}_K$ and $A_1$ are fitted to the energies of the first two states of the $K=1/2$ band, while the LEC $A_{\rm rot}$ is still fitted to the rotor.
Starting at N$^{2}$LO, all the LECs in the energy formula are fitted to the energies of the lowest states of the $K=1/2$ band.
Table \ref{table:LECs} lists the LECs employed at each order to describe the $K=1/2$ bands considered in this work.

\subsection{Poster children:\\ $\boldsymbol{K=1/2}$ bands in $\boldsymbol{^{167}}$Tm and $\boldsymbol{^{169}}$Er}

\begin{figure*}[htb!]
    \centering
    \includegraphics[width=\textwidth]{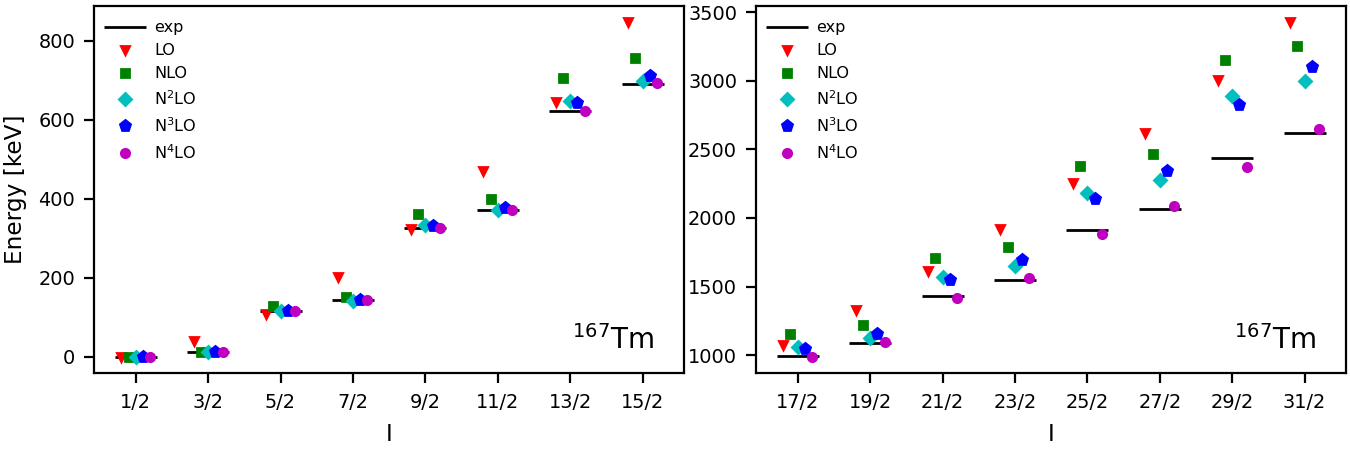}
    \caption{Calculated energy for the 1/2$^+$ ground-state rotational band in $^{167}$Tm. The black line shows the experimental values taken from the NNDC \cite{Baglin:2000ong}. The red triangles, green squares, cyan diamonds, blue pentagons and magenta circles are the calculated energies at LO, NLO, N$^2$LO, N$^3$LO, and N$^4$LO respectively. The right panel is a continuation of the left panel with a different scale for the y-axis.}
    \label{fig:E_167Tm}
\end{figure*}

\begin{figure}[b]
    \centering
    \includegraphics[width=\linewidth]{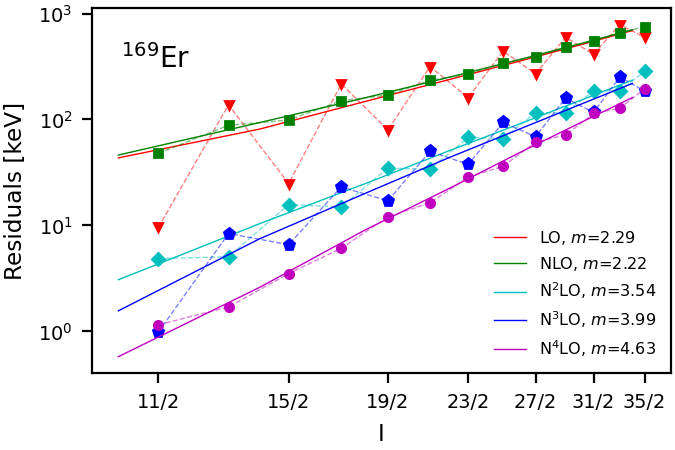}
    \caption{Energy residuals for the 1/2$^-$ ground-state rotational band in $^{169}$Er on a log-log scale. The red triangles, green squares, cyan diamonds, blue pentagons and magenta circles are the residuals from the calculated energies at LO, NLO, N$^2$LO, N$^3$LO, and N$^4$LO respectively. The dashed transparent lines are there to guide the eye. The solid lines show the trend of the calculated residuals after averaging out the signature staggering. The slope shown in the legend is the slope of the solid lines.}
    \label{fig:L_169Er}
\end{figure}

\begin{figure}[b]
    \centering
    \includegraphics[width=\linewidth]{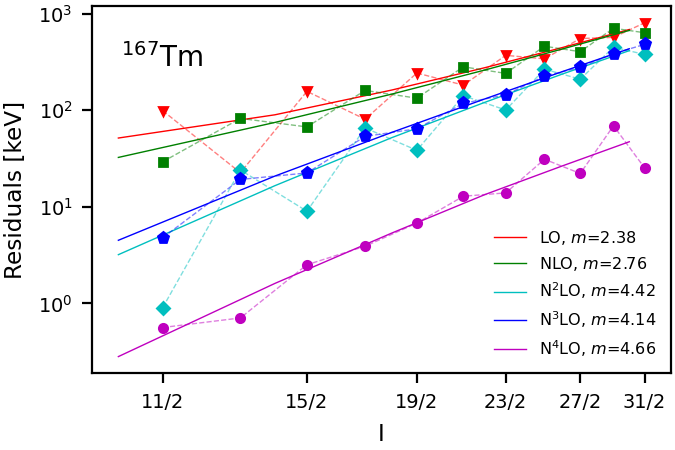}
    \caption{Energy residuals for the 1/2$^+$ ground-state rotational band in $^{167}$Tm on a log-log scale. The red triangles, green squares, cyan diamonds, blue pentagons and magenta circles are the residuals from the calculated energies at LO, NLO, N$^2$LO, N$^3$LO, and N$^4$LO respectively. The dashed transparent lines are there to guide the eye. The solid lines show the trend of the calculated residuals after averaging out the signature staggering. The slope shown in the legend is the slope of the solid lines.}
    \label{fig:L_167Tm}
\end{figure}

For the description of the 1/2$^-$ and 1/2$^+$ ground-state rotational bands of ${}^{169}$Er and ${}^{167}$Tm we use ${}^{168}$Er and ${}^{166}$Er as rotors, respectively.
Figures~\ref{fig:E_169Er}
%\begin{figure*}[h!]
%    \centering
%    \includegraphics[width=\textwidth]{Rotor Fermion/Figures/energy_169Er.png}
%    \caption{Calculated energies for the 1/2$^-$ ground-state rotational band in $^{169}$Er. The black line shows experimental values taken from the NNDC. Red triangles, green squares, cyan diamonds, blue pentagons and magenta circles show calculated energies at LO, NLO, N$^2$LO, N$^3$LO, and N$^4$LO, respectively. The right panel is a continuation of the left panel with a different scale for the y-axis.}
%    \label{fig:E_169Er}
%\end{figure*}
and~\ref{fig:E_167Tm}
show the calculated energies of these bands up to N$^4$LO together with experimental data.
In Figures~\ref{fig:L_169Er}
and~\ref{fig:L_167Tm}
the absolute residuals between theory and experiment, $|E_{\rm theo} - E_{\rm exp}|$, are plotted as a function of the total angular momentum of the system, $I$, on a log-log plot.
To gain insight as to how the error in our calculations scales with $I$, we remove the staggering of the absolute residuals, clearly seen in the log-log plots, by averaging the residuals of each pair of neighboring levels. This yields the solid lines in Figs.~\ref{fig:L_169Er} and~\ref{fig:L_167Tm}. If the error scales with a definite power of $I$, as expected in our EFT, these averaged residuals should follow a straight line with a slope greater than or equal to that power in the log-log plots. The slope of the line that best fits the averaged residuals is given in these figures' legends.

\begin{figure*}[t]
    \centering
    \includegraphics[width=\textwidth]{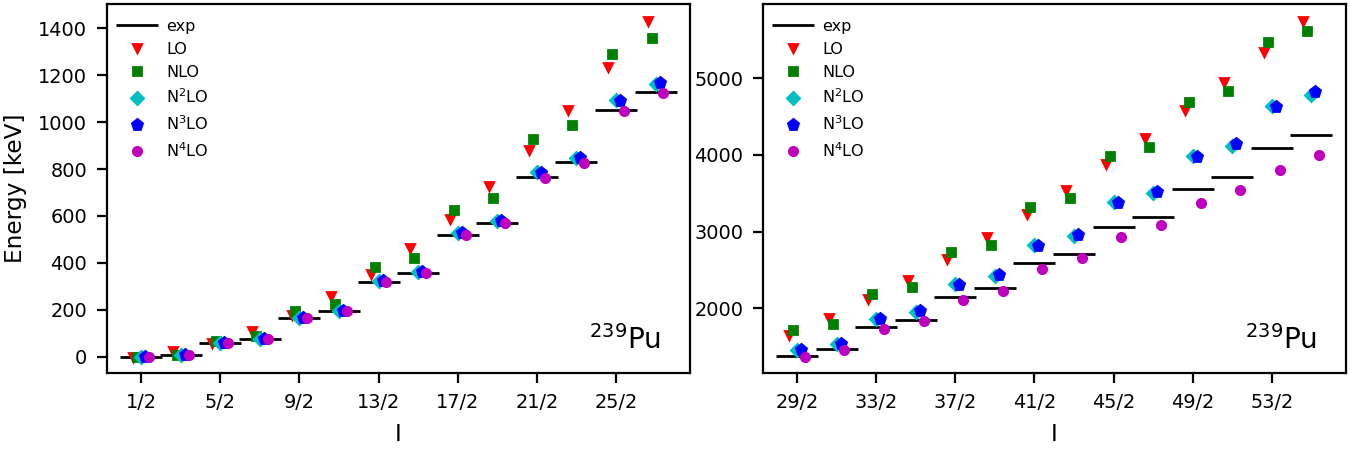}
    \caption{Calculated energy for the 1/2$^+$ ground-state rotational band in $^{239}$Pu. The black line shows the experimental values taken from the NNDC \cite{Browne:2014gwf}. The red triangles, green squares, cyan diamonds, blue pentagons and magenta circles are the calculated energies at LO, NLO, N$^2$LO, N$^3$LO, and N$^4$LO respectively. The right panel is a continuation of the left panel with a different scale for the y-axis.}
    \label{fig:E_239Pu}
\end{figure*}

For $^{169}$Er and $^{167}$Tm, the log-log plots make evident the systematic improvement of the calculated energies. Going from LO to NLO removes the energy staggering seen at LO (red triangles). Refitting the moment of inertia at N$^2$LO yields a clear improvement over NLO, as it permits the removal of errors of order $I^2$. This increases the slope of the averaged residuals significantly and gives us better agreement with experiment. However, N$^2$LO calculations reveal that there is staggering at higher orders which cannot be removed by the (NLO) signature term, proportional to $I$. Adding the correction to this term proportional to $I^3$ at N$^3$LO, removes most of this higher-order staggering. In $^{167}$Tm, the N$^3$LO energy formula gives us better qualitative agreement with experiment. However, the averaged error increases slightly as signaled by the decrease in the slope of the averaged residuals. For $^{169}$Er, the staggering is not clearly removed at N$^3$LO but the slope of the average residuals increases. Finally adding the N$^4$LO correction to the energy formula, proportional to $I^4$, improves the agreement with experiment dramatically, increasing the slope of the averaged residuals. We therefore see systematic order-by-order improvement across the known rotational levels of the ground-state bands of $^{169}$Er and $^{167}$Tm. The increasing slopes of the residuals make evident that the theory will eventually break down at higher $I$. Indeed, we already almost see this breakdown at the highest known levels.

There is similar systematic improvement for ${}^{169}$Tm but fewer levels so we do not discuss this case in the main text. Plots for that case which correspond to Figs.~\ref{fig:E_167Tm} and \ref{fig:L_167Tm} are provided in Appendix~\ref{ap:othernuclei}. 
${}^{167}$Er based on a ${}^{166}$Er core behaves similarly to ${}^{169}$Er, but there is less staggering, so we also relegate it to Appendix~\ref{ap:othernuclei}. The similarity of the results for ${}^{169}$Tm and ${}^{167}$Er to the cases presented in this section is not surprising given that the energy scales in all four systems are very similar.

\subsection{More complicated, yet still successful cases:\\$\boldsymbol{K=1/2}$ bands in $\boldsymbol{^{239}}$Pu and $\boldsymbol{^{235}}$U}

\begin{figure}[htb!]
    \centering
    \includegraphics[width=\linewidth]{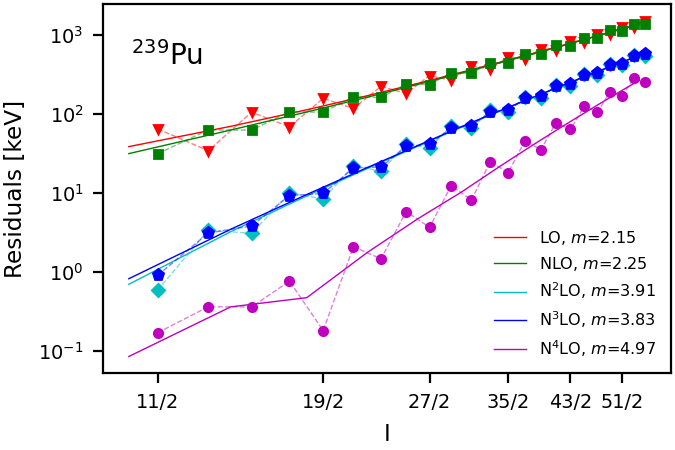}
    \caption{Energy residuals for the 1/2$^+$ ground-state rotational band in $^{239}$Pu on a log-log scale. The red triangles, green squares, cyan diamonds, blue pentagons and magenta circles are the residuals from the calculated energies at LO, NLO, N$^2$LO, N$^3$LO, and N$^4$LO respectively. The dashed transparent lines are there to guide the eye. The solid lines show the trend of the calculated residuals after averaging out the signature staggering. The slope shown in the legend is the slope of the solid lines.}
    \label{fig:L_239Pu}
\end{figure}

The case of $^{239}$Pu is more complicated as there are more single-particle levels close together.
A variant of the EFT presented in this work was already successfully applied to $^{239}$Pu  in Ref.~\cite{Papenbrock:2020zhh}. There $^{238}$Pu was chosen as the rotor, and we make the same choice. 
We study the rotational band built on the 1/2$+$ ground state, and carry out the analysis up to N$^4$LO. Figure~\ref{fig:E_239Pu} shows a clear systematic improvement in the agreement with data as we go to higher orders. Moving to the residuals, Fig.~\ref{fig:L_239Pu} shows good order-by-order improvement both in the size of residuals and in removal of the energy staggering. The slopes of the lines that best fit the averaged residuals in this plot increase as expected going from NLO to N$^2$LO and from N$^3$LO to N$^4$LO.

Papenbrock and Weidenm\"uller~\cite{Papenbrock:2020zhh} chose the energy of the $5/2^+$ bandhead at 300 keV as the breakdown scale in $^{239}$Pu. %This is the first nonrotational state above the 1/2$^+$ band that we---and they---consider.
We instead take $E_{\rm sp}=752$ keV since this is the energy of the first bandhead above the $1/2^+$ band with $|\Delta K| \leq 1$ and so sets the scale for mixing with the $1/2^+$ at N$^2$LO. We find the scales of the LECs are consistent with our power counting and this choice of $E_{\rm sp}$. $\Delta A$ could be considered an exception to this statement, but this somewhat large shift in $A$ at N$^2$LO can be understood in terms of the Nilsson model. There we expect the fermion to be in a large $j$ orbital which increases the size of the Coriolis coupling.

\begin{figure*}[t]
    \centering
    \includegraphics[width=\textwidth]{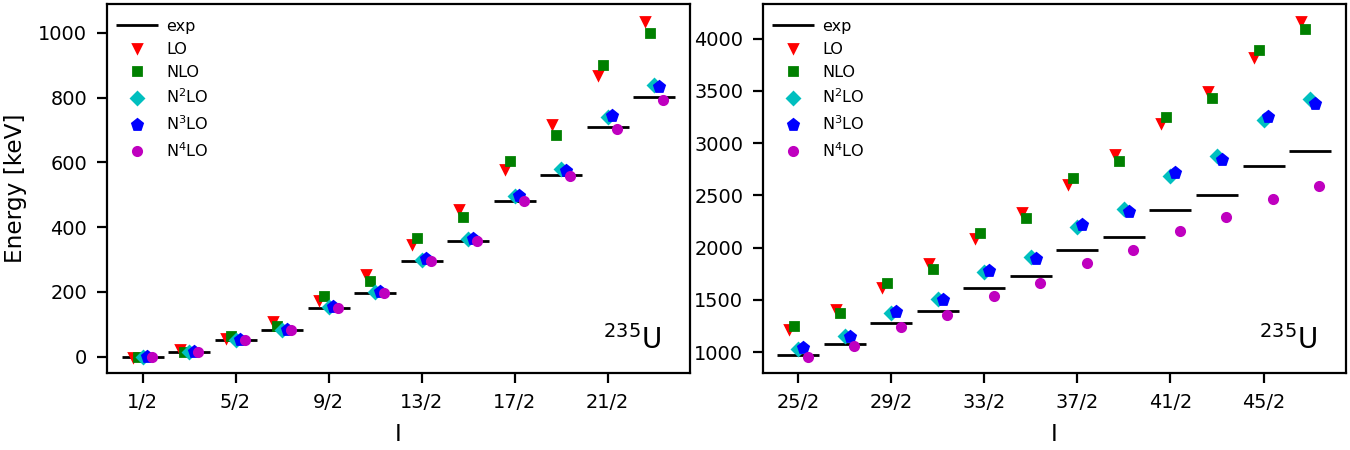}
    \caption{Calculated energy for states in the 1/2$^+$ excited-state rotational band in $^{235}$U. The black line shows the experimental values taken from the NNDC \cite{Browne:2014ukl}. The red triangles, green squares, cyan diamonds, blue pentagons and magenta circles are the calculated energies at LO, NLO, N$^2$LO, N$^3$LO, and N$^4$LO respectively. The right panel is a continuation of the left panel with a different scale on the y-axis.}
    \label{fig:E_235U}
\end{figure*}

\begin{figure}[b]
    \centering
    \includegraphics[width=\linewidth]{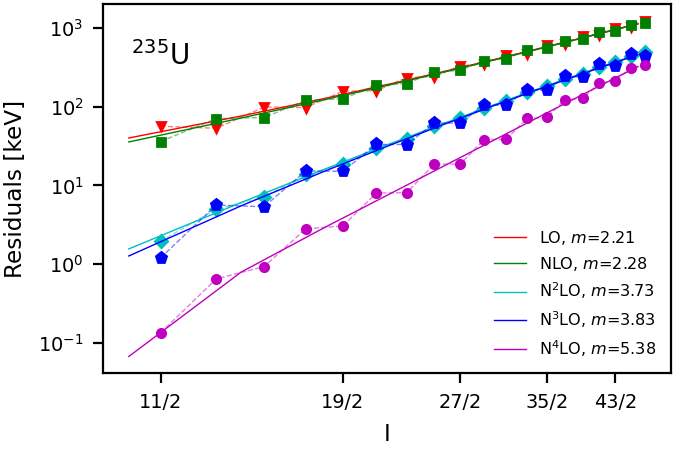}
    \caption{Energy residuals for states in the 1/2$^+$ excited-state rotational band in $^{235}$U on a log-log scale. The red triangles, green squares, cyan diamonds, blue pentagons and magenta circles are the residuals from the calculated energies at LO, NLO, N$^2$LO, N$^3$LO, and N$^4$LO respectively. The dashed transparent lines are there to guide the eye. The solid lines show the trend of the calculated residuals after averaging out the signature staggering. The slope shown in the legend is the slope of the solid lines.}
    \label{fig:L_235U}
\end{figure}

For ${}^{235}$U we see that our expansion works well, similarly to $^{239}$Pu. In this case we study the rotational band built on top of the first excited $1/2^+$ state of ${}^{235}$U and consider ${}^{234}$U to be our rotor. We focus on this $1/2^+$ band here because the band whose bandhead is the $7/2^-$ ground state of ${}^{235}$U couples to many nearby negative-parity bands. This leads to a complicated situation with several closely-spaced energy scales, see Ref.~\cite{BohrMottelson}. 
But, for the rotational band built on the selected $1/2^+$ state, systematic improvement of the agreement between theory and experiment is clearly seen in Figures~\ref{fig:E_235U} and~\ref{fig:L_235U}.

\subsection{$\boldsymbol{K=3/2}$ bands too: $\boldsymbol{^{159}}$Dy}

We can also use our EFT result to describe $K=3/2$ bands. As with $K=1/2$ bands, we employ the lowest states in the considered band to fit the parameter $E_K$ in the energy formula at LO. At NLO there is no correction to the enegry of the $K=3/2$ band.  At N$^2$LO and beyond, $K=3/2$ bands are described in terms of the rotor's effective moment of inertia and the energy of the bandhead. At N$^2$LO band-dependent terms shift the LEC $A_{\rm rot}$, effectively changing the moment of inertia. At N$^3$LO the signature term proportional to $I^3$ produces the dominant energy staggering in these bands. That staggering is typically less pronounced than that observed in $K=1/2$ bands, in agreement with our power counting.

\begin{figure*}[t]
    \centering
    \includegraphics[width=\textwidth]{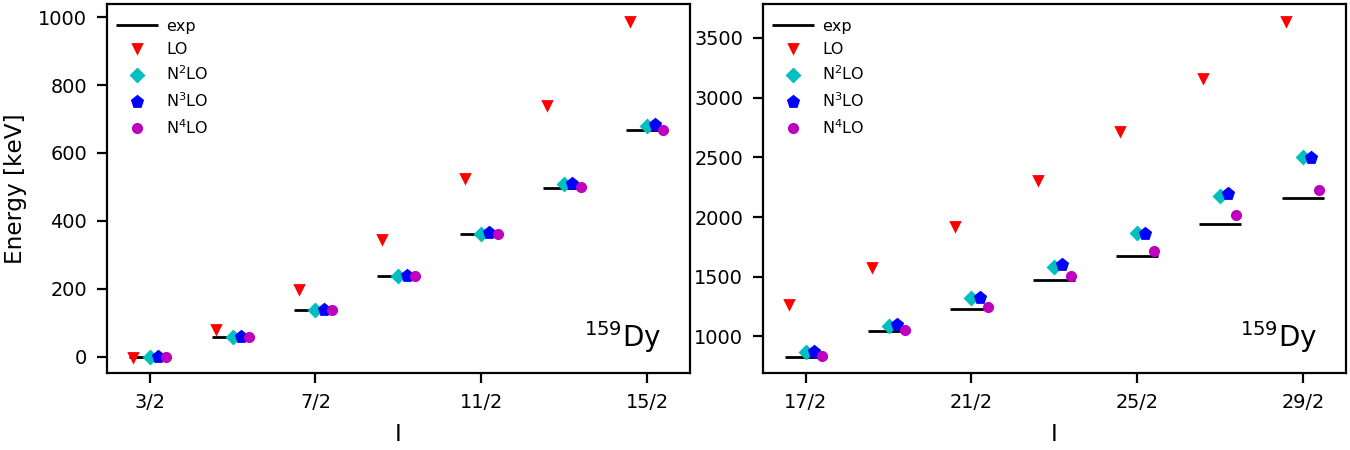}
    \caption{Calculated energies for states in the 3/2$^-$ ground-state rotational band in $^{159}$Dy. The black lines show the experimental values taken from the NNDC \cite{Reich:2012ouk}. The red triangles, cyan diamonds, blue pentagons and magenta circles are the calculated energies at LO, N$^2$LO, N$^3$LO, and N$^4$LO respectively. The right panels are a continuation of the left panels with a different scale on the y-axis.}
    \label{fig:E_3/2}
\end{figure*}

\begin{figure}[b]
    \centering
    \includegraphics[width=\linewidth]{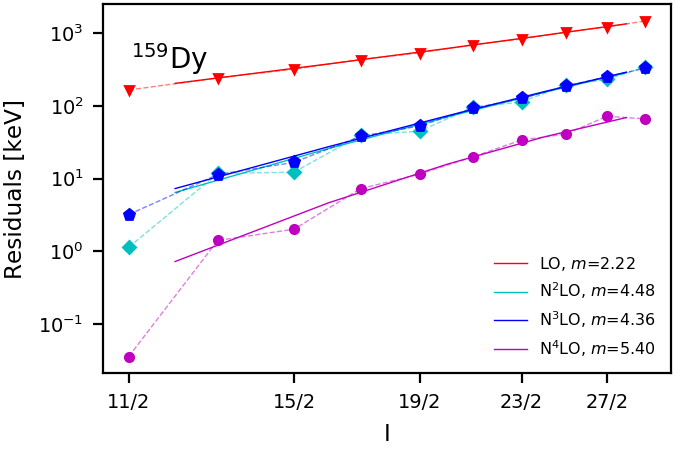}
    \caption{Energy residuals for the 3/2$^-$ ground-state rotational band in $^{159}$Dy on a log-log scale. The red triangles, cyan diamonds, blue pentagons and magenta circles are the residuals from the calculated energies at LO, N$^2$LO, N$^3$LO, and N$^4$LO respectively. The dashed transparent lines are there to guide the eye. The solid lines show the trend of the calculated residuals after averaging out the signature staggering. The slope shown in the legend is the slope of the solid lines.}
    \label{fig:L_3/2}
\end{figure}

To assess the EFT's performance for $K=3/2$ bands we need a case where there is a significant amount of data on the band's energy levels, and where other bands for which $|\Delta K|   \leq 1$ are separated by appreciable energy gaps from the band of interest. $^{159}$Dy, where the ground state has $I=3/2$, provides such a case. For this band there is clear systematic improvement as shown in Fig.~\ref{fig:E_3/2} and Fig.~\ref{fig:L_3/2}. From LO to N$^2$LO the slope of the residuals in the log-log plot increases by more than two units. From N$^2$LO to N$^3$LO the energy staggering (proportional to $I^3$) is almost completely removed. Finally, at N$^4$LO the slope improves to $5$--$6$, consistent with the idea that it is $I^5$ staggering and an $I^6$ term that are the dominant omitted effects.

%In ${}^{235}$U we use our EFT to describe the band on top of the $3/2^+$ state at 393.2 keV. After the N$^4$LO correction is included the absolute residuals become extremely small: $1$ keV or less until well beyond $I=10$. Additionally, this band does not have data on the energy staggering beyond the $I=11/2^+$-/$13/2^+$ pair. The combination of this data gap and the very small N$^4$LO residuals makes it difficult to discern a trend at that order. The left panel of Fig.~\ref{fig:L_3/2} makes it clear, though, that there is systematic improvement from LO to N$^2$LO and from N$^2$LO to N$^3$LO. Including the staggering term at N$^3$LO also decreases the size of the residuals, i.e., removes most of the staggering in the three pairs where we have information on it. Overall, the description of the studied $K=3/2$ band improves order-by-order, but does not show as clear systematic behavior as the results for the $K=1/2$ band in ${}^{235}$U described above do.
\begin{figure*}[htb!]
    \centering
    \includegraphics[width=\textwidth]{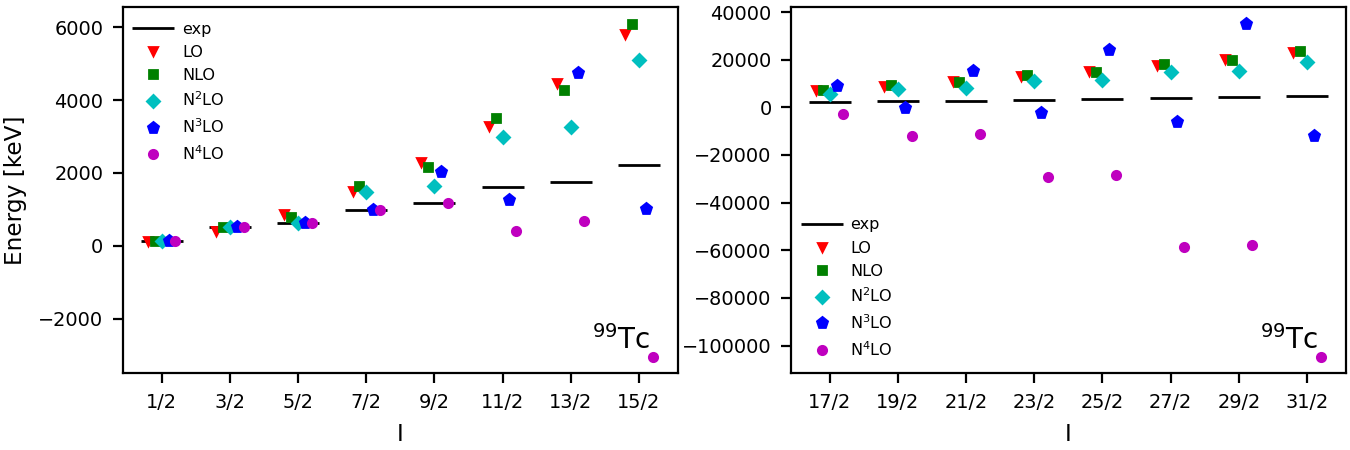}
    \includegraphics[width=\textwidth]{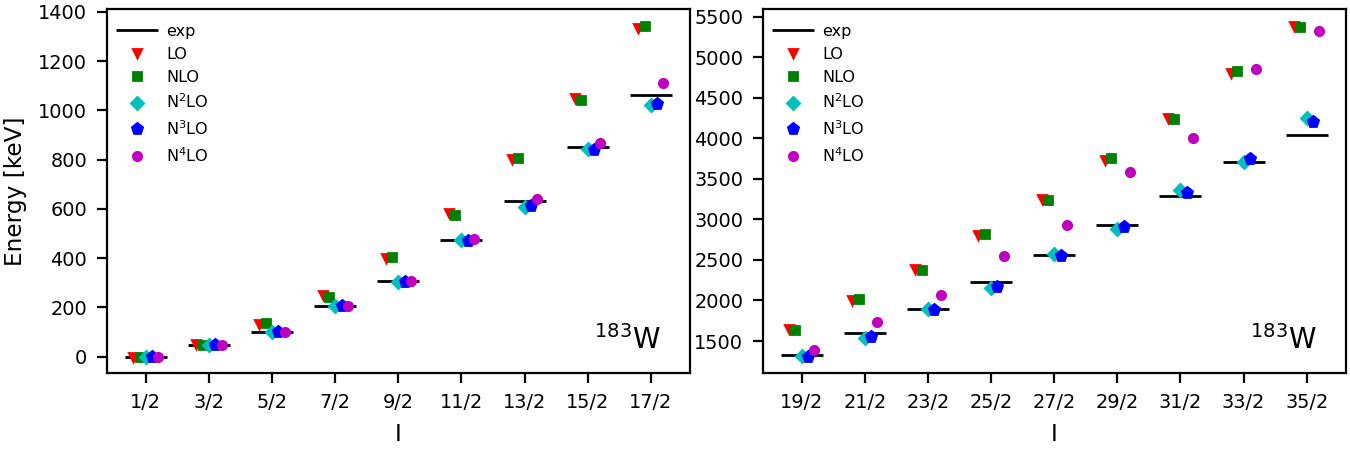}
    \caption{Calculated energy for states in the 1/2$^-$ excited-state rotational band in ${}^{99}$Tc (top panels) and in the 1/2$^-$ ground-state rotational band in ${}^{183}$W (bottom panels). The black line shows the experimental values taken from the NNDC \cite{Browne:2017uto,Baglin:2016vll}. The red triangles, green squares, cyan diamonds, blue pentagons and magenta circles are the calculated energies at LO, NLO, N$^2$LO, N$^3$LO, and N$^4$LO respectively. The right panel is a continuation of the left panel with a different scale for the y-axis.}
    \label{fig:E_99Tc}
\end{figure*}

\subsection{What failure looks like: $\boldsymbol{{}^{99}}$Tc}

\begin{figure*}[htb!]
    \centering
    \includegraphics[width=0.495\linewidth]{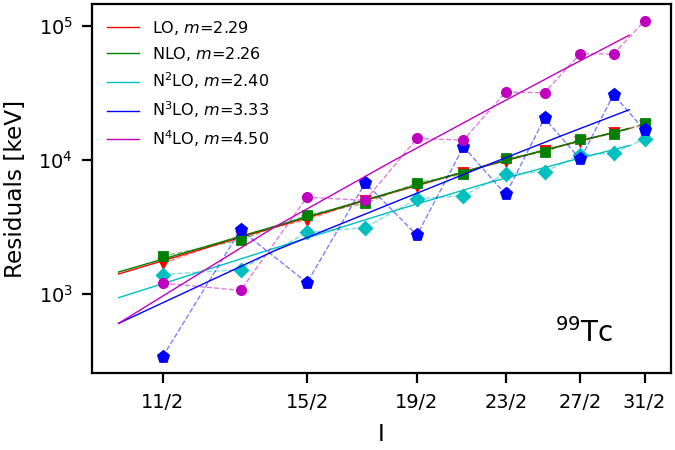}
    \includegraphics[width=0.495\linewidth]{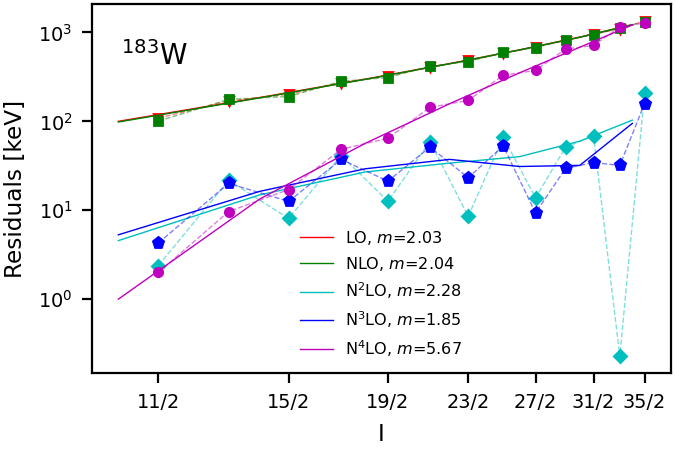}
    \caption{Energy residuals for states in the 1/2$^-$ excited-state rotational band in ${}^{99}$Tc (top panels) and in the 1/2$^-$ ground-state rotational band in ${}^{183}$W (bottom panels) on a log-log scale. The red triangles, green squares, cyan diamonds, blue pentagons and magenta circles are the residuals from the calculated energies at LO, NLO, N$^2$LO, N$^3$LO, and N$^4$LO respectively. The dashed transparent lines are there to guide the eye. The solid lines show the trend of the calculated residuals after averaging out the signature staggering. The slope shown in the legend is the slope of the solid lines.}
    \label{fig:L_99Tc}
\end{figure*}

For ${}^{99}$Tc we look at the rotational band built on top of the first 1/2$^-$ excited state. We consider ${}^{99}$Tc to be a proton hole on top of $^{100}$Ru as the rotor. We expect the breakdown scale for ${}^{99}$Tc to be very low since $\epsilon_{\rm sp}$ is greater than 1. We clearly see this in the top panels in Fig.~\ref{fig:E_99Tc} and the left panel in Fig.~\ref{fig:L_99Tc}, where going to higher order doesn't necessarily describe the data better. In fact, at N$^4$LO the theory prediction does worse than the predictions at lower orders when we go beyond the 17/2$^-$. Indeed, apart from the levels used in the fit, we could describe the energies of all levels better at lower orders. We also do not see the expected increase in the slope going from NLO to N$^2$LO. The magenta line crossing all the other lines in figure \ref{fig:L_99Tc} at low energies is a quantitative measure of the low breakdown energy of the fermion-rotor EFT in this case.

%\begin{figure*}[t]
%    \centering
%    \includegraphics[width=\textwidth]{Rotor Fermion/Figures/energy_183W.png}
%    \caption{Calculated energy for the 1/2$^-$ ground-state rotational band in $^{183}$W. The black line shows the experimental values taken from the NNDC. The red triangles, green squares, cyan diamonds, blue pentagons and magenta circles are the calculated energies at LO, NLO, N$^2$LO, N$^3$LO, and N$^4$LO respectively. The right panel is a continuation of the left panel with a different scale for the y-axis.}
%    \label{fig:E_183W}
%\end{figure*}

%\begin{figure}[b]
%    \centering
%    \includegraphics[width=\linewidth]{Rotor Fermion/Figures/log-log_183W.png}
%    \caption{Energy residuals for the 1/2$^-$ ground-state rotational band in $^{183}$W on a log-log scale. The red triangles, green squares, cyan diamonds, blue pentagons and magenta circles are the residuals from the calculated energies at LO, NLO, N$^2$LO, N$^3$LO, and N$^4$LO respectively. The dashed transparent lines are there to guide the eye. The solid lines show the trend of the calculated residuals after averaging out the signature staggering. The slope shown in the legend is the slope of the solid lines.}
%    \label{fig:L_183W}
%\end{figure}

\subsection{Two nearby bands: $\boldsymbol{{}^{183}}$W}

\label{sec:W183}

For $^{183}$W we study the 1/2$^-$ ground-state band and take the LEC $A$ from $^{182}$W as our rotor. We use the same procedure we used previously to get our LECs at each order. 

The bottom panels of Fig.~\ref{fig:E_99Tc} show the calculated energies for the ground-state band of $^{183}$W at all orders together with experimental data. We have to note that for $^{183}$W, there exists a low-lying 3/2$^-$ state at 208.8 keV which makes $\epsilon_{\rm sp}=0.48$. This makes our expansion parameter larger than the usual cases and we expect our EFT to break down relatively quickly. This could be understood in terms of band mixing between the 1/2$^-$ and 3/2$^-$ bands \cite{Rowe}. A way to fix this would be to include the 3/2$^-$ band as an additional low-energy degree of freedom in the Lagrangian. This would require us to fit the two bands simultaneously and we would then expect to get better agreement with data. 

This kind of EFT treatment of the rotor-fermion problem was implemented by Papenbrock and Weidenm\"uller for ${}^{187}$Os~\cite{Papenbrock:2020zhh}. They argued that this provides an EFT definition of triaxiality. Conversely, for rotational states whose excitation energies are less than $E_{\rm sp}$ one can can still perturb around the axial limit.

\begin{table*}[t]
\centering
\begin{tabular}{|c|cc||ccc||ccc||ccc||ccc|}
\hline
\multirow{2}{*}{Nucleus}&\multicolumn{2}{c||}{Bandhead}&\multicolumn{3}{c||}{Energy scales [keV]}&\multicolumn{3}{c||}{N$^2$LO}&\multicolumn{3}{c||}{N$^3$LO}&\multicolumn{3}{c|}{N$^4$LO}\\
&$E$ [keV] & $J^{\pi}$ & $E_{ \rm rot}$ & $E_{ \rm vib}$ & $E_{ \rm sp}$& ${\Delta A}/{A}$ & $\epsilon_{\rm sp}$ & $\epsilon_{\rm vib}$ & ${B_1}/{A}$  & $(\epsilon_{\rm sp})^2$ & $(\epsilon_{\rm vib})^2$ & ${B}/{A}$  & $(\epsilon_{\rm sp})^3$ & $(\epsilon_{\rm vib})^2$  \\\hline
$^{99}$\textbf{Tc} & 143 &1/2$^-$&         539   & 1362 & 529 & 0.21 & 1.02 & 0.40 & -0.1192 & 1.039 & 0.157 & -0.0216 & 1.0590 & 0.157 \\
$^{167}$\textbf{Er}&208  &1/2$^-$&         80     & 785  & 545 & 0.17 & 0.15 & 0.10 & -0.0019 & 0.022 & 0.010 & -0.0007 & 0.0032 & 0.010 \\
$^{169}$\textbf{Er}&0    &1/2$^-$&         80     & 821  & 562 & 0.12 & 0.14 & 0.10 & -0.0013 & 0.020 & 0.009 & -0.0004 & 0.0029 & 0.009 \\
$^{167}$\textbf{Tm}&0    &1/2$^+$&         80     & 785  & 470 & 0.08 & 0.17 & 0.10 & 0.0018  & 0.029 & 0.010 & -0.0007 & 0.0049 & 0.010 \\
$^{169}$\textbf{Tm}&0    &1/2$^+$&         80     & 821  & 570 & 0.07 & 0.14 & 0.10 & 0.0013  & 0.020 & 0.009 & -0.0004 & 0.0027 & 0.009 \\
$^{183}$\textbf{W} &0    &1/2$^-$&          100  & 1221 & 209 & 0.22 & 0.48 & 0.08 & -0.0006 & 0.230 & 0.007 & 0.0010  & 0.1102 & 0.007 \\
$^{235}$\textbf{U} &0.076&1/2$^+$&         43     & 927  & 393 & 0.17 & 0.11 & 0.05 & -0.0005 & 0.012 & 0.002 & -0.0004 & 0.0014 & 0.002 \\
$^{239}$\textbf{Pu}&0    &1/2$^+$&         44     & 605  & 752 & 0.15 & 0.06 & 0.07 & 0.0002  & 0.003 & 0.005 & -0.0001 & 0.0002 & 0.005 \\  \hline        
&$E$ [keV]& $J^{\pi}$ & $E_{ \rm rot}$ & $E_{ \rm vib}$ & $E_{ \rm sp}$& ${\Delta A}/{A}$ & $\epsilon_{\rm sp}$ & $\epsilon_{\rm vib}$ & ${A_3}/{A}$  & $(\epsilon_{\rm sp})^2$ & $(\epsilon_{\rm vib})^2$ & ${B}/{A}$  & $(\epsilon_{\rm sp})^3$ & $(\epsilon_{\rm vib})^2$  \\\hline
$^{159}$\textbf{Dy}&0    &3/2$^-$&  99 & 990 & 310 & 0.31 & 0.32 & 0.10 & -0.0003 & 0.102 & 0.010 & -0.0003 & 0.032 & 0.010 \\
%$^{235}$\textbf{U} &393  &3/2$^+$& 43 & 927 & 393 & 0.07 &&&-0.0002 &&&-0.0003 &&\\
 \hline 
\end{tabular}
\caption{In the first block of the table we show the rotational bandhead energy, spin and parity. In the second block energy scales for different nuclei are shown in units of keV. In the third block we compare the relative correction to $A$ at N$2$LO, ${\Delta A}/{A}$, to its expected size $\epsilon_{\rm sp},\epsilon_{\rm vib}$. In the next block, ${B_1}/{A}$ (or $A_3/A$ for $3/2$ bands) is the relative correction to the energy at N$^3$LO; its expected size is $(\epsilon_{\rm sp},\epsilon_{\rm vib})^2$. Finally, at N$^4$LO the relative correction is ${B}/{A}$ and this is expected to be of order $(\epsilon_{\rm sp})^3$ or$(\epsilon_{\rm vib})^2$. The top block of the table reports results for $K=1/2$ bands and the bottom block gives results for $K=3/2$ bands.}
\label{table:1}
\end{table*}

The staggering in $^{183}$W in is not clearly present in experimental data and therefore we do not see a clear improvement going from LO to NLO. Going to N$^2$LO we see a clear improvement overall and at N$^4$LO we only see improvement for the levels with low $I$. This comes from the relatively large expansion parameter and is consistent with our expectation that the EFT breaks down relatively early.

 The right panel of Fig.~\ref{fig:L_99Tc} shows the log-log plot of the residuals where we clearly see the breakdown at around $I=15/2$, where the N$^2$LO and N$^3$LO lines cross the N$^4$LO line. The very low residual at $I \approx 15$ for N$^2$LO is accidental: the residuals shift from being negative to being positive there. This accidental crossing also explains the bending of the N$^2$LO and N$^3$LO lines for $I > 10$.

\subsection{Values and order-by-order stability of LECs}

\label{sec:LECs}

We show the bandhead properties, relevant energy scales, and the relative sizes of LECs for the systems studied in this work in Table~\ref{table:1}. The third, fourth, and fifth segments of the table show the relative size of the LEC that appears at that order compared to the LO LEC $A$. Each block then compares that relative size to the expectation based on energy-scale ratios in the nucleus of interest. We note that it is sometimes hard to decide whether $\epsilon_{\rm sp}$ or $\epsilon_{\rm vib}$ sets the size of the correction at each order and indeed, one sometimes sees an interplay between both. The ratios fall in the expected range except for a few cases. For $^{239}$Pu ($^{235}${U}) we see that ${\Delta A}/{A}$ is two times (1.5 times) larger than both $\epsilon_{\rm sp}$ and $\epsilon_{\rm vib}$. This is consistent with natural coefficients in the EFT expansion. It could be related to the large Coriolis coupling associated with high $j$ orbitals for the fermion. $^{239}$Pu and $^{235}${U} are large nuclei and we expect the intrinsic wave functions for both nuclei to have sizable intrinsic angular momentum for the last nucleon. We also notice two nuclei where $E_{\rm sp}$ is comparable to $E_{\rm rot}$, $^{99}${Tc} and $^{183}${W}. In those two cases we expect the breakdown scale to be very low and our EFT not to be very useful.

\begin{table*}[hbt]
\centering
\begin{tabular}{|c|ccccc||cccc||cc||c|}
\hline
\multirow{2}{*}{Nucleus} & \multicolumn{5}{c||}{$A$ {[}keV{]}} & \multicolumn{4}{c||}{$A_1$ {[}keV{]}}         & \multicolumn{2}{c||}{$B_1$ {[}keV{]}} & $B$ {[}keV{]} \\
                         & LO    & NLO   & N$^2$LO  & N$^3$LO  & N$^4$LO  & NLO     & N$^2$LO   & N$^3$LO  & N$^4$LO  & N$^3$LO            & N$^4$LO           & N$^4$LO        \\\hline
$^{99}$\textbf{Tc}       & 89.92 & 89.92 & 71.40 & 57.65 & 79.02 & 32.22   & 50.74  & 83.39 & 60.31 & -6.874          & -3.454         & -1.710      \\
$^{167}$\textbf{Er}      & 13.43 & 13.43 & 11.18 & 11.14 & 11.24 & 5.59    & 7.84   & 7.94  & 7.83  & -0.021          & -0.005         & -0.008      \\
$^{169}$\textbf{Er}      & 13.30 & 13.30 & 11.76 & 11.73 & 11.78 & 8.22    & 9.75   & 9.83  & 9.77  & -0.015          & -0.007         & -0.004      \\
$^{167}$\textbf{Tm}      & 13.43 & 13.43 & 12.35 & 12.40 & 12.50 & -9.96   & -8.88  & -8.99 & -9.10 & 0.023           & 0.039          & -0.008      \\
$^{169}$\textbf{Tm}      & 13.30 & 13.30 & 12.38 & 12.41 & 12.47 & -10.50  & -9.58  & -9.65 & -9.72 & 0.016           & 0.026          & -0.005      \\
$^{183}$\textbf{W}       & 16.68 & 16.68 & 13.01 & 12.99 & 12.84 & -1.19   & 2.49   & 2.52  & 2.69  & -0.008          & -0.033         & 0.013       \\
$^{235}$\textbf{U}       & 7.25  & 7.25  & 6.03  & 6.02  & 6.05  & -2.93   & -1.71  & -1.69 & -1.73 & -0.003          & 0.002          & -0.003      \\
$^{239}$\textbf{Pu}      & 7.34  & 7.34  & 6.25  & 6.25  & 6.27  & -4.72   & -3.63  & -3.64 & -3.66 & 0.001           & 0.004          & -0.002     \\    \hline
\end{tabular}
\caption{The different LECs at each order for $K=1/2$ bands. Note that going from NLO to N$^2$LO we do not add a new LEC, however at LO and NLO we fit $A$ to the the rotational band in the rotor, while at N$^2$LO and beyond we fit $A$ to the band in the odd-mass nucleus.}
\label{table:LECs}
\end{table*}

In Table~\ref{table:LECs} we show, for each of the nuclei whose $K=1/2$ bands we have studied in this work, the values of the LECs obtained at each order. For the nuclei where we have a good separation of scales we see that the LECs are stable going from order to order. Since we take $A$ from the rotor at LO and at NLO, it only changes at N$^2$LO where we re-fit the moment of inertia. The size of change for $A$ is consistent with the power counting, as shown in Table~\ref{table:1} and discussed in the previous paragraph.

\begin{table}[hbt]
\centering
\begin{tabular}{|c|cccc||cc||c|}
\hline
\multirow{2}{*}{Nucleus} & \multicolumn{4}{c||}{$A$ {[}keV{]}} & \multicolumn{2}{c||}{$A_3$ {[}keV{]}} & {$B$ {[}keV{]}} \\
                         & LO      & N$^2$LO  & N$^3$LO  & N$^4$LO & N$^3$LO            & N$^4$LO           & N$^4$LO        \\\hline
${}^{159}$\textbf{Dy} & 16.48 & 11.32 & 11.35 & 11.44 & -0.005 & -0.009 & -0.005 \\ 
%${}^{235}$\textbf{U} & 7.25 & 6.70 & 6.71 & 6.75 & -0.001 & -0.003 & -0.002 \\ 
\hline
\end{tabular}
\caption{The different LECs at each order for $K=3/2$ bands. Note that going from LO to N$^2$LO we do not add a new LEC. However at LO we fit $A$ to the the rotational band in the rotor, while at N$^2$LO and beyond we fit $A$ to the band in the odd-mass nucleus.}
\label{table:3/2bandLECs}
\end{table}
The $A_1$ parameter also changes when going from NLO to N$^2$LO and then stabilizes for nuclei with good separation of scales. The sizable change is because we remove the $I^2$ dependence when we fit the $A$ parameter to the odd mass system at N$^2$LO. Until that happens $A_1$ is contaminated by $\sim I^2$ effects. For $^{183}$W we see a large shift in $A_1$ going from NLO to N$^2$LO, but this is because there is no clear staggering in the experimental data for this nucleus. $^{99}$Tc has the largest fluctuations in all its LECs because of the large expansion parameter $\epsilon_{\rm sp} >1$. 

We note that $B$ is negative for most nuclei. This can be understood from the rigid rotor model where when we allow for small fluctuations around rigidity the moment of inertia increases with increasing $I$, due to centrifugal distortions.

Table \ref{table:3/2bandLECs} provides the LECs for the nucleus where we examined data on a $K=3/2$ band: ${}^{159}$Dy. As in the studied $K=1/2$ bands, the change in $A$ from LO to N$^2$LO is consistent with expectations from the power counting. The values for this LEC at higher orders do not change drastically. The variation of the LEC that enters at N$^3$LO and accompanies the $K=3/2$ signature term, $A_3$, with the EFT order is similar to that of $B_1$.

\subsection{Extracting the breakdown scale in different systems}

We extract the breakdown scale by locating where the N$^4$LO line (magenta) crosses the lower-order lines in the log-log plots. That is to say: we define the theory to have broken down when the theory prediction at N$^3$LO does better that the prediction at N$^4$LO. This occurs within the range of the plot for $^{99}$Tc and $^{183}$W, where we identify the breakdown scale to be at 1500 keV and 820 keV respectively. We note that these numbers are higher than the naive breakdown scale associated with other single-particle energies. We see a similar higher-than-expected breakdown scale for the other nuclei considered. In fact, for all but ${}^{99}$Tc and ${}^{183}$W, the N$^4$LO line never crosses the N$^3$LO line within the domain of levels considered in this study, even though we go well beyond the single-particle and vibrational energy scales.

\section{Summary and Outlook}

\label{sec:conclusions}

In their 1969 book Bohr and Mottelson give a formula for the energies of rotational bands in odd-mass nuclei and explain how that formula arises from the particle-rotor model. In this work we have shown how this description of  rotational bands in odd-mass nuclei that are built on a single fermion level can be understood as an effective field theory (EFT). The expansion parameter in the EFT is $v$, the rotational velocity of the system. The expansion in $v$ in the Lagrangian becomes a dual expansion in powers of $\epsilon_{\rm sp}$ and $\epsilon_{\rm vib}$ in the Hamiltonian and for the band's energy levels. We worked out the energy-level formula to fourth order in this expansion and fitted the parameters therein for the systems ${}^{167,169}$Tm, ${}^{167,169}$Er, ${}^{239}$Pu, ${}^{235}$U, ${}^{159}$Dy, ${}^{99}$Tc, and ${}^{183}$W. The EFT gives a good description of rotational energy bands to surprisingly high spin in the first seven cases, but fails in the last two. 

We showed that this EFT viewpoint on rotational bands in odd-mass nuclei can help to explain why the particle-rotor model works where it does and predict its accuracy in a particular system. Through analysis of both the EFT's low-energy constants and its order-by-order residuals we showed for these nine systems that the size of different effects in the energy-level formula is in line with the power counting in the EFT. The EFT's accuracy in a particular nucleus is connected to the underlying energy scales there.

Our study therefore goes beyond the related discussion of an EFT for rotational bands in odd-mass nuclei by Papenbrock and Weidenm\"uller in Ref.~\cite{Papenbrock:2020zhh}. That work considered effects up to N$^2$LO in $v$, and discussed only two different nuclides, ${}^{239}$Pu and ${}^{187}$Os. It also did not perform an order-by-order analysis of residuals with respect to data to demonstrate systematic improvement. Moreover, our EFT has a conceptual difference to that of Ref.~\cite{Papenbrock:2020zhh}. Papenbrock and Weidenm\"uller implicitly assumed that operators in the EFT could also be organized in powers of the fermionic degrees of freedom $K$, $\vec{j}$, and $\vec{r}$. Here we make no such assumption.

That is because we want our results to be independent of the model of the underlying nuclear dynamics. In a particular model of fermionic orbitals, e.g., the Nillson model, some of the low-energy constants appearing in our EFT's Hamiltonian could be predicted. While we agree that they can be estimated, 
we argue that there are too many unknowns for any particular single-particle model to give a reliable prediction for the EFT coefficients. And indeed, there is a long and not particularly successful string of efforts to explain in the particle-rotor picture why the Coriolis coupling tends to be overestimated once a specific model for the single-particle orbitals is adopted (see, e.g., Ref.~\cite{Hamamoto:2011wn} for a summary). 
The most conservative path forward is thus to fit the EFT's formula to data. Connecting the coefficients in the rotational-band formula to the underlying dynamics could be an interesting subject for future work although it should be noted that an incredible amount of effort has been spent in this direction in the past. 

A straightforward next step now that we have an EFT Lagrangian that is a good description of fermionic rotational bands is to include electromagnetic fields and compute intra-band transitions. In even-even nuclei that step generates parameter-free predictions at leading-order accuracy~\cite{CoelloPerez:2015}. Another avenue for future work is to use Bayesian parameter estimation to obtain the parameters in the EFT at each order~\cite{Schindler:2009,Wesolowski:2016,Wesolowski:2019}. In this work the EFT's coefficients were obtained using the lowest energy levels and assuming no theoretical uncertainty. The Bayesian methodology of Refs.~\cite{Schindler:2009,Wesolowski:2016,Wesolowski:2019} ensures that EFT parameters are stable as more orders are included in the fit, because it includes the effects of higher-order terms on those parameters. Finally, we point out that a longer-term goal is to apply this EFT to halo nuclei in which low-lying rotational states of the core play a prominent role, such as ${}^{11}$Be and ${}^{31}$Ne. In such an application the fermionic dynamics---or at least part of it---could be explicitly calculated in Halo EFT~\cite{Hammer:2017}.

\acknowledgments{DRP is grateful for the warm hospitality of the IKP Theoriezentrum Darmstadt, where part of this work was carried out. We thank Mark Caprio and Thomas Papenbrock for useful discussions.
This work was supported by the US Department of Energy, contract DE-FG02-93ER40756 (IKA, DRP), by the ExtreMe Matter Institute (DRP), by the Deutsche Forschungsgesellschaft under Grant SFB 1245 (EACP), and under the auspices of the US Department of Energy by Lawrence Livermore National Laboratory under Contract DE-AC52-07NA27344 (EACP). IKA Acknowledges the support of King Saud University and the Ministry of Education in Saudi Arabia.}

\appendix
\begin{widetext}
\section{Expectation Value of the NLO Hamiltonian}
\label{app:1}
The expectation value of the shift in the Hamiltonian at NLO is
%\begin{eqnarray}
%&& \langle{KIM}|\Delta \hat{H}_{\rm NLO}|{KIM}\rangle\nonumber\\
%&& \qquad =\frac{C_1}{C_0}\langle{KIM}|(\hat{J}_{+1}\hat{Q}_{-1}+\hat{J}_{-1}\hat{Q}_{+1})|{KIM}\rangle\nonumber\\
%&& \qquad =\frac{2C_1}{C_0}\langle{KIM}|\hat{J}_{+1}\hat{Q}_{-1}|{KIM}\rangle
%\end{eqnarray}
\begin{equation}
\begin{aligned}
\langle{KIM}|\Delta \hat{H}_{\rm NLO}|{KIM}\rangle
=& \frac{C_1}{C_0}\langle{KIM}|(\hat{J}_{+1}\hat{Q}_{-1}+\hat{J}_{-1}\hat{Q}_{+1})|{KIM}\rangle\\
=& \frac{2C_1}{C_0}\langle{KIM}|\hat{J}_{+1}\hat{Q}_{-1}|{KIM}\rangle
\end{aligned}
\end{equation}
Applying $\hat{Q}_{-1}$ to $\Psi_{KIM}$ gives us
%\begin{eqnarray}
%\label{A2}
%&& \hat{Q}_{-1}\Psi_{KIM}\nonumber\\
%&& =\frac{\hbar }{\sqrt{2}} \left(\frac{2I+1}{16\pi^2}\right)^{1/2}
% \Big(\xi_K\sqrt{I(I+1)-K(K+1)}\mathscr{D}^{I}_{M(K+1)}\nonumber\\
%&&  \qquad +(-1)^{I+K}\xi_{\bar{K}}\sqrt{I(I+1)+K(-K+1)}\mathscr{D}^{I}_{M(-K+1)}\Big).\nonumber\\
%\end{eqnarray}
%\begin{widetext}
\begin{equation}
\hat{Q}_{-1}\Psi_{KIM}
=\frac{\hbar}{\sqrt{2}} \sqrt{\frac{2I+1}{16\pi^2}}
 \left[\xi_K\sqrt{I(I+1)-K(K+1)}\mathscr{D}^{I}_{M(K+1)}
+(-1)^{I+K}\xi_{\bar{K}}\sqrt{I(I+1)+K(-K+1)}\mathscr{D}^{I}_{M(-K+1)}\right]
\label{A2}
\end{equation}
%\end{widetext}
Then we have
%\begin{eqnarray}
%&&\langle{KIM}|\Delta \hat{H}_{\rm NLO}|{KIM}\rangle=\nonumber\\
%&&  \frac{2\hbar C_1}{\sqrt{2}C_0}\left(\frac{2I+1}{16\pi^2}\right) \int d\omega\Big(\xi^*_K\mathscr{D}^{*I}_{MK}+(-1)^{I+K}\xi^*_{\bar{K}}\mathscr{D}^{*I}_{M-K}\Big)\nonumber\\
%&&  \qquad(\hat{J}_{+1})\Big(\xi_K\sqrt{I(I+1)-K(K+1)}\mathscr{D}^{I}_{M(K+1)}\nonumber\\
%&&  \qquad +(-1)^{I+K}\xi_{\bar{K}}\sqrt{I(I+1)+K(-K+1)}\mathscr{D}^{I}_{M(-K+1)}\Big)\nonumber\\
%&&  =\frac{2\hbar  C_1}{\sqrt{2}C_0}(-1)^{I+K}\left(\frac{2I+1}{16\pi^2}\right) \xi^*_K \hat{J}_{+1}\xi_{\bar{K}} \nonumber\\
%&& \qquad \int d\omega \mathscr{D}^{*I}_{MK}\sqrt{I(I+1)+K(-K+1)}\mathscr{D}^{I}_{M(-K+1)}
%\end{eqnarray}
\begin{equation}
\begin{aligned}
\langle{KIM}|\Delta \hat{H}_{\rm NLO}|{KIM}\rangle =&
\frac{2 C_1}{C_0}\sqrt{\frac{2I+1}{16\pi^2}} \int d\Omega\left[\xi^*_K\mathscr{D}^{*I}_{MK}+(-1)^{I+K}\xi^*_{\bar{K}}\mathscr{D}^{*I}_{M-K}\right]
\hat{J}_{+1}\hat{Q}_{-1}\Psi_{KIM}\\
=&\frac{2\hbar  C_1}{\sqrt{2}C_0}(-1)^{I+K}\left(\frac{2I+1}{16\pi^2}\right) \xi^*_K \hat{J}_{+1}\xi_{\bar{K}}
\int d\omega \mathscr{D}^{*I}_{MK}\sqrt{I(I+1)+K(-K+1)}\mathscr{D}^{I}_{M(-K+1)}
\end{aligned}
\end{equation}
The last line is non-zero only when $K=1/2$ and this gives us
%\begin{eqnarray}
%&& \langle{1/2IM}|\Delta \hat{H}_{\rm NLO}|{1/2IM}\rangle\nonumber\\
%&& \qquad =\frac{\hbar C_1}{2C_0}(-1)^{I+1/2}(I+1/2)\langle{1/2}|\sqrt{2}\hat{J}_{+1}|{\overline{1/2}}\rangle.\nonumber\\
%\end{eqnarray}
\begin{equation}
\langle{1/2IM}|\Delta \hat{H}_{\rm NLO}|{1/2IM}\rangle\nonumber
=\frac{\hbar C_1}{2C_0}(-1)^{I+1/2}(I+1/2)\langle{1/2}|\sqrt{2}\hat{J}_{+1}|{\overline{1/2}}\rangle.
\end{equation}

\section{N$\boldsymbol{^2}$LO Matrix Elements}
\label{app:2}
We want to calculate
%\begin{eqnarray}
%&& \langle KIM| (\hat{J}_{+1}\hat{Q}_{-1}+\hat{J}_{-1}\hat{Q}_{+1})^2|KIM\rangle\nonumber\\
%&& \qquad =\sum_\nu |\langle \nu IM| (\hat{J}_{+1}\hat{Q}_{-1}+\hat{J}_{-1}\hat{Q}_{+1})|KIM\rangle|^2%\nonumber\\
%&& \qquad =4\sum_\nu |\langle \nu IM|\hat{J}_{+1}\hat{Q}_{-1}|KIM\rangle|^2
%\end{eqnarray}
\begin{equation}
\langle KIM| (\hat{J}_{+1}\hat{Q}_{-1}+\hat{J}_{-1}\hat{Q}_{+1})^2|KIM\rangle
=\sum_\nu |\langle \nu IM| (\hat{J}_{+1}\hat{Q}_{-1}+\hat{J}_{-1}\hat{Q}_{+1})|KIM\rangle|^2.
\end{equation}
From equation~(\ref{A2}) and integrating over the Wigner D-matrices we have the following matrix elements
%\begin{widetext}
%\begin{eqnarray}
%&&\langle{\nu IM}|\hat{J}_{+1}\hat{Q}_{-1}|{KIM}\rangle\nonumber\\
%&& \qquad =\frac{\hbar}{4}\Bigg[\delta_{\nu,K+1}\sqrt{I(I+1)-K(K+1)}\langle{\nu}|\sqrt{2} \hat{J}_{+1}|{K}\rangle \nonumber\\
%&& \qquad \quad+(-1)^{I+K}\delta_{\nu,-K+1}\sqrt{I(I+1)-K(K-1)}\langle{\nu}|\sqrt{2} \hat{J}_{+1}|{\bar{K}}\rangle\nonumber\\
%&& \qquad\quad +(-1)^{I+\nu}\delta_{\nu,-K-1}\sqrt{I(I+1)-K(K+1)}\langle{\bar{\nu}}|\sqrt{2} \hat{J}_{+1}|{K}\rangle\nonumber\\
%&& \qquad \quad +(-1)^{2I+K+\nu}\delta_{\nu,K-1}\sqrt{I(I+1)-K(K-1)}\langle{\bar{\nu}}|\sqrt{2}\hat{J}_{+1}|{\bar{K}}\rangle\Bigg]
%\end{eqnarray}
\begin{equation}
\begin{aligned}
\langle{\nu IM}|\hat{J}_{+1}\hat{Q}_{-1}|{KIM}\rangle
= \frac{\hbar}{4}\Bigg[&\delta_{\nu,K+1}\sqrt{I(I+1)-K(K+1)}\langle{\nu}|\sqrt{2} \hat{J}_{+1}|{K}\rangle\\
&+(-1)^{I+K}\delta_{\nu,-K+1}\sqrt{I(I+1)-K(K-1)}\langle{\nu}|\sqrt{2} \hat{J}_{+1}|{\bar{K}}\rangle\\
&+(-1)^{I+\nu}\delta_{\nu,-K-1}\sqrt{I(I+1)-K(K+1)}\langle{\bar{\nu}}|\sqrt{2} \hat{J}_{+1}|{K}\rangle\nonumber\\
&+(-1)^{2I+K+\nu}\delta_{\nu,K-1}\sqrt{I(I+1)-K(K-1)}\langle{\bar{\nu}}|\sqrt{2}\hat{J}_{+1}|{\bar{K}}\rangle\Bigg]
\end{aligned}
\end{equation}
We also have
%\begin{eqnarray}
%&&\langle{\nu IM}|\hat{J}_{-1}\hat{Q}_{+1}|{KIM}\rangle\nonumber\\
%&& \qquad =\frac{\hbar}{4}\Bigg[\delta_{\nu,K-1}\sqrt{I(I+1)-K(K-1)}\langle{\nu}|\sqrt{2}\hat{J}_{-1}|{K}\rangle\nonumber\\
%&& \qquad\quad +(-1)^{I+K}\delta_{\nu,-K-1}\sqrt{I(I+1)-K(K+1)}\langle{\nu}| \sqrt{2}\hat{J}_{-1}|{\bar{K}}\rangle\nonumber\\
%&& \qquad\quad +(-1)^{I+\nu}\delta_{\nu,-K+1}\sqrt{I(I+1)-K(K-1)}\langle{\bar{\nu}}|\sqrt{2}\hat{J}_{-1}|{K}\rangle\nonumber\\
%&& \qquad\quad +(-1)^{2I+K+\nu}\delta_{\nu,K+1}\sqrt{I(I+1)-K(K+1)}\langle{\bar{\nu}}|\sqrt{2}\hat{J}_{-1}|{\bar{K}}\rangle\Bigg]
%\end{eqnarray}
\begin{equation}
\begin{aligned}
\langle{\nu IM}|\hat{J}_{-1}\hat{Q}_{+1}|{KIM}\rangle
=\frac{\hbar}{4}\Bigg[&\delta_{\nu,K-1}\sqrt{I(I+1)-K(K-1)}\langle{\nu}|\sqrt{2}\hat{J}_{-1}|{K}\rangle\\
&+(-1)^{I+K}\delta_{\nu,-K-1}\sqrt{I(I+1)-K(K+1)}\langle{\nu}| \sqrt{2}\hat{J}_{-1}|{\bar{K}}\rangle\\
&+(-1)^{I+\nu}\delta_{\nu,-K+1}\sqrt{I(I+1)-K(K-1)}\langle{\bar{\nu}}|\sqrt{2}\hat{J}_{-1}|{K}\rangle\\
&+(-1)^{2I+K+\nu}\delta_{\nu,K+1}\sqrt{I(I+1)-K(K+1)}\langle{\bar{\nu}}|\sqrt{2}\hat{J}_{-1}|{\bar{K}}\rangle\Bigg]
\end{aligned}
\end{equation}
%For $\nu=1/2$ and $K=1/2$ we get
%\begin{eqnarray}
%&&\langle{1/2IM}|\hat{J}_{+1}\hat{Q}_{-1}|{1/2IM}\rangle\nonumber\\
%&& \qquad=\frac{\hbar}{4}(-1)^{I+K} \sqrt{I(I+1)+K(-K+1)}\langle{1/2}| \sqrt{2}\hat{J}_{+1}|{\overline{1/2}}\rangle
%\end{eqnarray}
%For $\nu=3/2$ and $K=1/2$ we get
%\begin{eqnarray}
%&&\langle{3/2IM}|\hat{J}_{+1}\hat{Q}_{-1}|{1/2IM}\rangle\nonumber\\
%&& \qquad=\frac{\hbar}{4}\delta_{\nu,K+1}\sqrt{I(I+1)-K(K+1)}\langle{3/2}| \sqrt{2}\hat{J}_{+1}|{1/2}\rangle
%\end{eqnarray}
This gives 
\begin{eqnarray}
\langle KIM| (\hat{J}_{+1}\hat{Q}_{-1}+\hat{J}_{-1}\hat{Q}_{+1})^2|KIM\rangle=a I(I+1) + b K^2 + c K
%\nonumber\\
%&& \qquad=\frac{\hbar^2}{4} \sum_{\nu=K\pm 1 or -\nu=K\pm 1}( I(I+1)-K(K\pm 1))\Big[|\langle{\nu}| \sqrt{2}\hat{J}_{\pm 1}|{{K}}\rangle|^2 +|\langle{\bar{\nu}}| \sqrt{2}\hat{J}_{\pm 1}|{K}\rangle|^2\Big]
\end{eqnarray}
where the coefficients $a$, $b$, and $c$ are comprised of the squares of matrix elements of $\hat{J}_{\pm 1}$ between the state $|K \rangle$ and states $| \nu \rangle$ and $|\bar{\nu} \rangle$ with $\nu=|K \pm 1|$.
%\end{widetext}

\section{Additional results: ${}^{169}$Tm and ${}^{167}$Er}
\label{ap:othernuclei}

\noindent
In Figs.~\ref{fig:Esupp} and~\ref{fig:Lsupp} we provide energy spectra order-by-order and log-log plots of residuals for ${}^{167}$Er and ${}^{169}$Tm. These are to be compared to the corresponding results in the main text for ${}^{169}$Er and ${}^{167}$Tm respectively. 

\begin{figure*}[htbp]
    \centering
    \includegraphics[width=\textwidth]{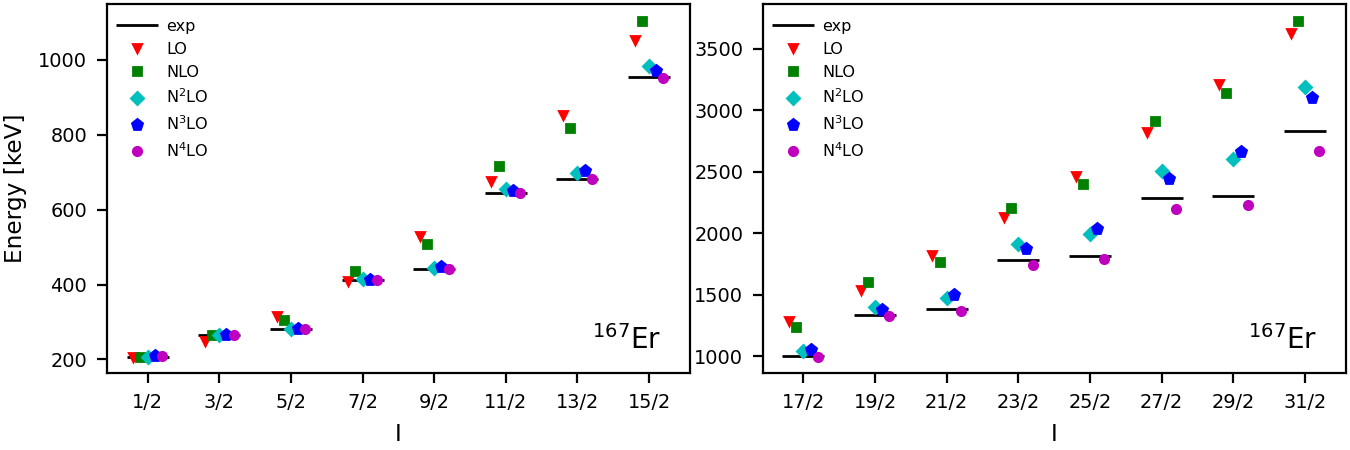}
    \includegraphics[width=\linewidth]{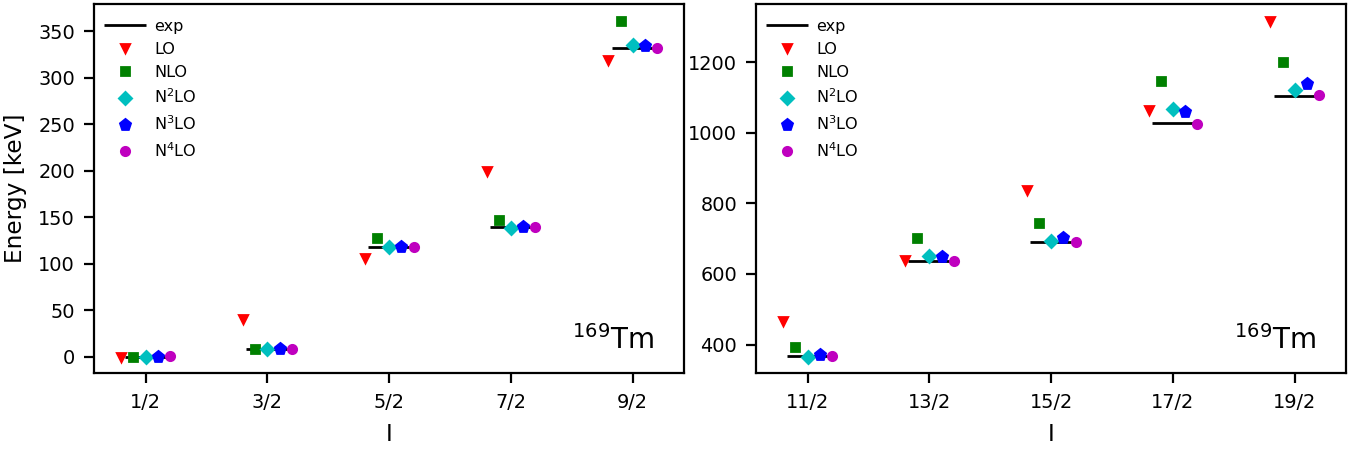}
    \caption{Calculated energy for states in the 1/2$^-$ excited-state rotational band in $^{167}$Er (top panels) and states in the 1/2$^+$ excited-state rotational band in $^{169}$Tm (bottom panels). The black line shows the experimental values taken from the NNDC \cite{Baglin:2000ong,Baglin:2008hsa}. The red triangles, green squares, cyan diamonds, blue pentagons and magenta circles are the calculated energies at LO, NLO, N$^2$LO, N$^3$LO, and N$^4$LO respectively. The right panel is a continuation of the left panel with a different scale for the y-axis.}
    \label{fig:Esupp}
\end{figure*}
\begin{figure*}[htbp]
    \centering
    \includegraphics[width=0.495\linewidth]{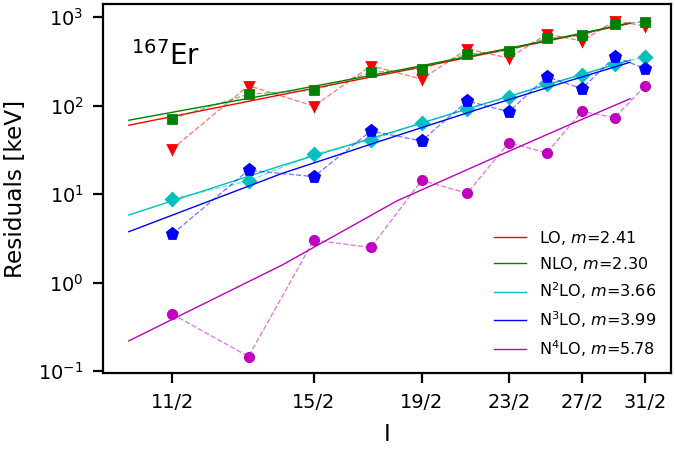}
        \includegraphics[width=0.495\linewidth]{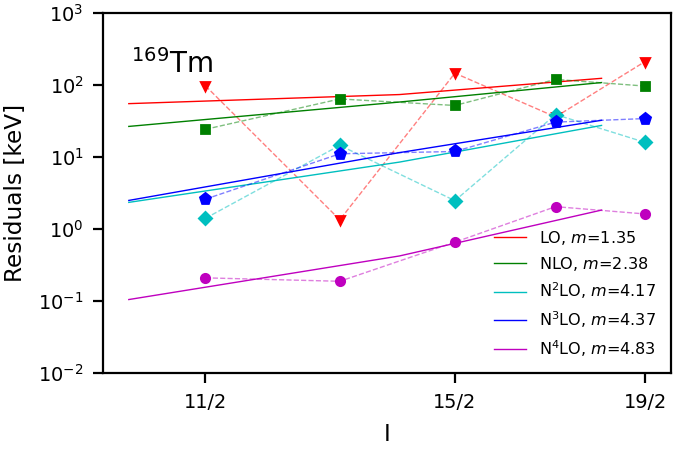}
    \caption{Energy residuals for the 1/2$^-$ excited-state rotational band in $^{167}$Er (left panel) and 1/2$^+$ ground-state rotational band in $^{169}$Tm (right panel) on a log-log scale. The red triangles, green squares, cyan diamonds, blue pentagons and magenta circles are the residuals from the calculated energies at LO, NLO, N$^2$LO, N$^3$LO, and N$^4$LO respectively. The dashed transparent lines are there to guide the eye. The solid lines show the trend of the calculated residuals after averaging out the signature staggering. The slope shown in the legend is the slope of the solid lines.}
    \label{fig:Lsupp}
\end{figure*}
\end{widetext}


\begin{thebibliography}{15}
\bibitem{BohrMottelson}
A.~Bohr and B.~Mottelson, Nuclear Structure, Volume II: Deformations (World Scientific, Singapore, 1998). 

\bibitem{Papenbrock:2010yg}
T.~Papenbrock,
%``Effective theory for deformed nuclei,''
Nucl. Phys. A \textbf{852}, 36 (2011)
doi:10.1016/j.nuclphysa.2010.12.013
[arXiv:1011.5026 [nucl-th]].
%23 citations counted in INSPIRE as of 08 Jul 2020}

\bibitem{CoelloPerez:2015}
   E.~A.~Coello~P\'erez and T.~Papenbrock,
    %``Effective theory for the non-rigid rotor in an electromagnetic field: Toward accurate and precise calculations of E2 transitions in deformed nuclei,''
  Phys.\ Rev.\ C {\bf 92}, no. 1, 014323 (2015)
  doi:10.1103/PhysRevC.92.014323
  [arXiv:1502.04405 [nucl-th]].
  %%CITATION = doi:10.1103/PhysRevC.92.014323;%%
  
  %\cite{Papenbrock:2020zhh}
\bibitem{Papenbrock:2020zhh}
T.~Papenbrock and H.~Weidenmüller,
%``Effective field theory for deformed odd-mass nuclei,''
[arXiv:2005.11865 [nucl-th]].
%0 citations counted in INSPIRE as of 01 Jun 2020

\bibitem{Rowe}
D.~J.~Rowe, Nuclear Collective Motion: Models and Theory, Chapter 6 (World Scientific, Singapore, 2010). 
  
  %\cite{Chen:2020qbf}
\bibitem{Chen:2020qbf}
Q.~Chen, N.~Kaiser, U.~G.~Meißner and J.~Meng,
%``Effective field theory for triaxially deformed odd-mass nuclei,''
[arXiv:2003.04065 [nucl-th]].
%1 citations counted in INSPIRE as of 01 Jun 2020

%\cite{Caprio:2019yxh}
\bibitem{Caprio:2019yxh}
M.~A.~Caprio, P.~J.~Fasano, P.~Maris, A.~E.~McCoy and J.~P.~Vary,
%``Probing ab initio emergence of nuclear rotation,''
Eur. Phys. J. A \textbf{56}, no.4, 120 (2020)
doi:10.1140/epja/s10050-020-00112-0
[arXiv:1912.00083 [nucl-th]].
%2 citations counted in INSPIRE as of 09 Jul 2020

\bibitem{Lepage:1997cs}
G.~P.~Lepage,
%``How to renormalize the Schrodinger equation,''
[arXiv:nucl-th/9706029 [nucl-th]].
%385 citations counted in INSPIRE as of 09 Jul 2020

%\cite{Morris:2017vxi}
\bibitem{Morris:2017vxi}
T.~D.~Morris, J.~Simonis, S.~R.~Stroberg, C.~Stumpf, G.~Hagen, J.~D.~Holt, G.~R.~Jansen, T.~Papenbrock, R.~Roth and A.~Schwenk,
%``Structure of the lightest tin isotopes,''
Phys. Rev. Lett. \textbf{120}, no.15, 152503 (2018)
doi:10.1103/PhysRevLett.120.152503
[arXiv:1709.02786 [nucl-th]].
%76 citations counted in INSPIRE as of 18 Jul 2020

%\cite{Rotureau:2016jpf}
\bibitem{Rotureau:2016jpf}
J.~Rotureau, P.~Danielewicz, G.~Hagen, F.~Nunes and T.~Papenbrock,
%``Optical potential from first principles,''
Phys. Rev. C \textbf{95}, no.2, 024315 (2017)
doi:10.1103/PhysRevC.95.024315
[arXiv:1611.04554 [nucl-th]].
%37 citations counted in INSPIRE as of 18 Jul 2020 

%\cite{Baglin:2008hsa}
\bibitem{Baglin:2008hsa}
C.~M.~Baglin,
%``Nuclear Data Sheets for A = 169,''
Nucl. Data Sheets \textbf{109}, no.9, 2033-2256 (2008)
doi:10.1016/j.nds.2008.08.001.
%14 citations counted in INSPIRE as of 02 Nov 2020

%\cite{Baglin:2000ong}
\bibitem{Baglin:2000ong}
C.~M.~Baglin,
%``Nuclear Data Sheets for A = 167,''
Nucl. Data Sheets \textbf{90}, no.3, 431-644 (2000)
doi:10.1006/ndsh.2000.0012
%11 citations counted in INSPIRE as of 02 Nov 2020

%\cite{Browne:2014gwf}
\bibitem{Browne:2014gwf}
E.~Browne and J.~K.~Tuli,
%``Nuclear Data Sheets for A = 239,''
Nucl. Data Sheets \textbf{122}, 293-376 (2014)
doi:10.1016/j.nds.2014.11.003
%4 citations counted in INSPIRE as of 02 Nov 2020

%\cite{Browne:2014ukl}
\bibitem{Browne:2014ukl}
E.~Browne and J.~K.~Tuli,
%``Nuclear Data Sheets for A = 235,''
Nucl. Data Sheets \textbf{122}, 205-292 (2014)
doi:10.1016/j.nds.2014.11.002
%7 citations counted in INSPIRE as of 02 Nov 2020

%\cite{Reich:2012ouk}
\bibitem{Reich:2012ouk}
C.~W.~Reich,
%``Nuclear Data Sheets for A = 159,''
Nucl. Data Sheets \textbf{113}, no.1, 157-363 (2012)
doi:10.1016/j.nds.2012.01.002
%12 citations counted in INSPIRE as of 02 Nov 2020

%\cite{Browne:2017uto}
\bibitem{Browne:2017uto}
E.~Browne and J.~K.~Tuli,
%``Nuclear Data Sheets for A = 99,''
Nucl. Data Sheets \textbf{145}, 25-340 (2017)
doi:10.1016/j.nds.2017.09.002
%7 citations counted in INSPIRE as of 02 Nov 2020

%\cite{Baglin:2016vll}
\bibitem{Baglin:2016vll}
C.~M.~Baglin,
%``Nuclear Data Sheets for A = 183,''
Nucl. Data Sheets \textbf{134}, 149-430 (2016)
doi:10.1016/j.nds.2016.04.002
%2 citations counted in INSPIRE as of 02 Nov 2020

%\cite{Schindler:2008fh}
\bibitem{Schindler:2009}
M.~R.~Schindler and D.~R.~Phillips,
%``Bayesian Methods for Parameter Estimation in Effective Field Theories,''
Annals Phys. \textbf{324}, 682-708 (2009)
doi:10.1016/j.aop.2008.09.003
[arXiv:0808.3643 [hep-ph]].

%\cite{Wesolowski:2015fqa}
\bibitem{Wesolowski:2016}
S.~Wesolowski, N.~Klco, R.~J.~Furnstahl, D.~R.~Phillips and A.~Thapaliya,
%``Bayesian parameter estimation for effective field theories,''
J. Phys. G \textbf{43}, no.7, 074001 (2016)
doi:10.1088/0954-3899/43/7/074001
[arXiv:1511.03618 [nucl-th]].
%60 citations counted in INSPIRE as of 24 Sep 2020

%\cite{Wesolowski:2018lzj}
\bibitem{Wesolowski:2019}
S.~Wesolowski, R.~J.~Furnstahl, J.~A.~Melendez and D.~R.~Phillips,
%``Exploring Bayesian parameter estimation for chiral effective field theory using nucleon–nucleon phase shifts,''
J. Phys. G \textbf{46}, no.4, 045102 (2019)
doi:10.1088/1361-6471/aaf5fc
[arXiv:1808.08211 [nucl-th]].
%35 citations counted in INSPIRE as of 24 Sep 2020

%\cite{Hammer:2017tjm}
\bibitem{Hammer:2017}
H.~W.~Hammer, C.~Ji and D.~R.~Phillips,
%``Effective field theory description of halo nuclei,''
J. Phys. G \textbf{44}, no.10, 103002 (2017)
doi:10.1088/1361-6471/aa83db
[arXiv:1702.08605 [nucl-th]].
%73 citations counted in INSPIRE as of 24 Sep 2020

%\cite{Hamamoto:2011wn}
\bibitem{Hamamoto:2011wn}
I.~Hamamoto and B.~Mottelson,
%``Shape Deformations in Atomic Nuclei,''
Scholarpedia \textbf{7}, no.4, 10693 (2011)
doi:10.4249/scholarpedia.10693
[arXiv:1107.5248 [nucl-th]].
%12 citations counted in INSPIRE as of 02 Nov 2020

\end{thebibliography}
\end{document}